\documentclass[12pt,letterpaper]{article}
\usepackage{times}
\usepackage{amssymb}
\usepackage{t1enc}
\usepackage[latin1]{inputenc}
\usepackage[english]{babel}
\usepackage{graphicx}
\renewcommand{\baselinestretch}{1.75}

\begin{document}

\title{Sums of Independent Circular Random Variables and Maximum Likelihood Circular Uniformity Tests Based on Nonnegative Trigonometric Sums Distributions}
\author{Fern\'andez-Dur\'an, J.J. and Gregorio-Dom\'inguez, M.M. \\
ITAM\\
%R\'io Hondo No. 1 \\
%Col. Progreso Tizap\'an \\
%C.P. 01080 \\
%CDMX \\
%M\'exico \\
%Telephone: (5255) 56 28 34 25 \\
E-mail: jfdez@itam.mx \\
Preprint: arXiv:2212.01416v2 [stat.ME]}
\renewcommand{\baselinestretch}{1.00}
\date{}
\maketitle

\newpage

\renewcommand{\baselinestretch}{1.75}

\begin{abstract}
The circular uniform distribution on the unit circle is closed under summation, that is, the sum of independent circular uniformly distributed random
variables is also circular uniformly distributed. In this study, it is shown that a family of circular distributions based on nonnegative trigonometric sums (NNTS) is
also closed
under summation. Given the flexibility of NNTS circular distributions to model multimodality and skewness, these are good candidates for use as alternative
models to test for circular uniformity to detect different deviations from the null hypothesis of circular uniformity. The circular uniform distribution is a member of
the NNTS family, but in the NNTS parameter space, it corresponds to a point on the boundary of the parameter space, implying that the regularity conditions are not satisfied when the parameters are estimated by using the maximum likelihood method. Two NNTS tests for circular uniformity were developed by considering the standardised maximum
likelihood estimator and the generalised likelihood ratio. Given the nonregularity condition, the critical values of the proposed NNTS circular uniformity tests were
obtained via simulation and interpolated for any sample size by the fitting of regression models. The validity of the proposed NNTS circular uniformity tests was evaluated by generating NNTS models close to the circular uniformity null hypothesis.
\end{abstract}


\newpage
%\textbf{Keywords}: .

%\renewcommand{\baselinestretch}{1.75}

\section{Introduction}

A circular random variable takes values on the unit circle, and its density function must be periodic. Generally, a circular random variable represents an angle
in the interval $(0,2\pi]$, and it is important to model many random phenomena in different areas, such as in meteorology for the wind direction, in biology for the
dihedral angles defining the spatial structure of a protein, in ecology for camera trap data that records the time of the day at which different animals are observed and in many other examples in other disciplines.

The circular distributions based on nonnegative trigonometric sums (NNTS) developed by Fern\'{a}ndez-Dur\'{a}n (2004) are flexible distributions
capable of modelling circular data that present multimodality and/or skewness (asymmetry). Fern\'andez-Dur\'{a}n and Gregorio-Dom\'{i}nguez (2010) developed an efficient
optimisation algorithm for manifolds to obtain maximum likelihood estimates of the parameters of an NNTS model. This algorithm was implemented by using the $R$ package
$CircNNTSR$ (Fern\'{a}ndez-Dur\'{a}n and Gregorio-Dom\'{i}nguez, 2016). The NNTS family of distributions includes the uniform distribution as a special case ($M$=0). One of the most important hypothesis tests in the study of circular statistics is testing for circular uniformity (see Fisher, 1993;
Mardia and Jupp,
2000). In addition, for absolutely continuous circular density functions, the circular uniform density is closed under summation, that is, the sum of independent circular uniformly
distributed random variables has a circular uniform distribution (see Mardia and Jupp, 2000). Additionally, if at least one member of the sum of independent circular random
variables is uniformly distributed, then the sum of these random variables is also circular uniformly distributed. This is a consequence of the characteristic
function of a circular uniformly distributed random variable, $\phi_{unif}(t)$, defined for $t \in \mathbb{R}$ as:
\begin{equation}
\phi_{unif}(t)=I(t=0)
\end{equation}
where $I(t=0)$ is an indicator function that takes the value of one for $t=0$ and zero otherwise.

There are many different tests for circular uniformity, such as the Rayleigh test, Watson's test (Watson, 1961), Kuyper's test (Kuiper, 1960), and Rao's spacing test (Rao,
1976),
among others. Many of these tests were designed with unimodal distributions as an alternative hypothesis and presented low power when applied to multimodal datasets.
The Hermans-Rasson test (Hermans and Rasson, 1985), Bogdan test (Bogdan et al., 2002) and Pycke test (Pycke, 2010) consider multimodal circular distributions as
alternative hypotheses and have more power to detect deviations from circular uniformity when applied to multimodal circular data. In this study, two tests are
developed with alternative
hypotheses NNTS distributions to account for the cases that present multimodality but also asymmetry. The NNTS tests are based on the maximum likelihood method:
The first test is based on the standardised maximum likelihood estimator, and the second is based on the generalised likelihood ratio statistic. The power of the NNTS tests
was compared to that of the Hermans-Rasson and Pycke tests. The results of the sums of NNTS random variables allow us to identify NNTS densities that are close to the
uniform distribution, and we use these results to compare the power of the tests in simulated datasets where the degree of closeness to the circular uniform distribution can be controlled.
This study is divided into seven sections, including the introduction. In the second section, mathematical formulas for the characteristic function of an NNTS
circular random variable are developed. In the third section, the NNTS family is shown to be closed under summation; that is, the sum of independent NNTS circular random variables is also NNTS
distributed, and how to obtain the parameters of the NNTS density of the sum and numerical examples with graphs is explained. In the fourth section, the two
proposed circular uniformity tests, taking an NNTS distribution as an alternative hypothesis are developed. Considering the parameter space of NNTS densities, the null
circular uniformity distribution corresponds to a parameter on the boundary of the parameter space. Then, the regularity conditions of maximum likelihood estimation are
not satisfied because the parameters of the NNTS densities are estimated by maximum likelihood, and the critical values of the NNTS circular uniformity test are obtained by
simulation. In the fifth section, the power of the NNTS circular uniformity tests is examined by considering the results of the sums of the NNTS random variables in the third
section to consider alternative circular distributions that are close to the null circular uniform distribution. The practical application of the proposed tests to real data
on the time of occurrence of earthquakes and the flying orientation of home pigeons is presented in the sixth section. Finally, in the seventh section, conclusions are presented.

\section{Characteristic Function of an NNTS Circular Random Variable}

A circular random variable $\Theta$ is defined as a random variable with unit circle support. These random variables are relevant when modelling
seasonal patterns in many different scientific areas.
Let $\Theta$ be a circular random variable with an NNTS distribution with support interval $[0,2\pi)$ with a density function defined as
the squared norm of a sum of complex trigonometric terms:
\begin{equation}
f_{\Theta}(\theta) = \frac{1}{2\pi}\left|\left|\sum_{k=0}^M c_k  e^{ik\theta}\right|\right|^2 = \frac{1}{2\pi}\sum_{k=0}^M\sum_{m=0}^M c_k\bar{c}_m e^{i(k-m)\theta}
\end{equation}
where $i=\sqrt{-1}$ and $e^{ik\theta}=\cos(k\theta) + i\sin(k\theta)$. Complex parameter vector $\underline{c}=(c_0, c_1, \ldots, c_M)$ with
complex numbers $c_k= c_{rk} + ic_{ik}$ and $\bar{c}_k= c_{rk} - ic_{ik}$, where $c_{rk}$ and $c_{ik}$ are the real and imaginary parts of the complex number $c_k$,
respectively. The parameter vector $\underline{c}$ must satisfy $\sum_{k=0}^M ||c_k||^2 = 1$
where $||c_k||^2=c_{rk}^2 + c_{ik}^2$ is the squared norm of the complex number $c_k$. Given this parameter constraint, $c_0$ must be a positive real number related to the density concentration around its modes. The parameter set is a subset of the surface of a complex unit hypersphere in the space of complex numbers of dimension
$M+1$, $\mathbb{C}^{M+1}$ because $\underline{c}$ and $-\underline{c}$ produce the same NNTS density. In addition, the vector of parameters $\underline{c}$ written in reverse
order, produces the same NNTS density. For identifiability, we considered the parameter vectors $\underline{c}$ with $c_0$ positive and $c_0^2 > ||c_{M+1}||^2$. The number of
terms in the sum $M$ is an additional parameter that determines the maximum number of modes of the NNTS density function. By increasing $M$, it is possible to increase
the number of modes and/or the concentration around the modes in the NNTS density function. The case $M=0$ corresponds to a circular uniform distribution,
$f_{\Theta}(\theta)=\frac{1}{2\pi}$. The NNTS density satisfies the periodicity constraint for a circular density $f(\theta)=f(\theta+(2\pi)r)$ for any integer
$r$.

The characteristic function of a circular random variable, $\phi_{\Theta}(t)$, is defined as
\begin{equation}
\phi_{\Theta}(t)=E\left(e^{it\theta}\right) = \int_{0}^{2\pi}e^{it\theta}f_{\Theta}(\theta)d\theta.
\end{equation}
The characteristic function of an NNTS circular random variable is obtained as
\begin{equation}
\phi_{\Theta}(t)=E\left(e^{it\theta}\right) = \int_{0}^{2\pi}e^{it\theta}\frac{1}{2\pi}\sum_{k=0}^M\sum_{m=0}^M c_k\bar{c}_m e^{i(k-m)\theta}d\theta =
\frac{1}{2\pi}\sum_{k=0}^M\sum_{m=0}^M c_k\bar{c}_m \int_{0}^{2\pi} e^{i(k-m+t)\theta}d\theta
\end{equation}
then
\begin{equation}
\phi_{\Theta}(t) = I(t=0) + \mathop{\sum_{k=0}^M\sum_{m=0}^M}_{k \ne m} c_k\bar{c}_m I(t=m-k)
\label{char1}
\end{equation}
where $I(t=a)$ is an indicator function that takes the value of one if $t=a$ and zero otherwise. This result is obtained because the integral $\int_{0}^{2\pi}e^{ir\theta}d\theta$ is zero for $r \ne 0$ and equal to $2\pi$ for $r=0$. Rearranging the terms in Equation (\ref{char1}), we obtain
\begin{equation}
\phi_{\Theta}(t) = I(t=0) + \sum_{k=-M}^{-1} \left(\sum_{j=0}^{M-k}c_j\bar{c}_{j+k}\right) I(t=k) + \sum_{k=1}^{M} \left(\sum_{j=0}^{M-k}c_{j+k}\bar{c}_{j}\right) I(t=k).
\end{equation}
Thus, the characteristic function of an NNTS circular random variable takes values on the integers $-M, -M+1, \ldots, -1, 0, 1, \ldots, M$, and these values are functions
of the vector of parameters $\underline{c}=(c_0, c_1, \ldots, c_M)^{\top}$.

\section{Distribution of the Sum of Independent NNTS Circular Random Variables}

Let $\Theta_1, \Theta_2, \ldots, \Theta_S$ be independent NNTS circular random variables with parameter vectors $\underline{c}^{(1)}$, $\underline{c}^{(2)}$, $\ldots$,
$\underline{c}^{(S)}$ and $M_1, M_2, \ldots, M_S$, respectively. For independent random variables, the characteristic function of $\sum_{k=1}^S \Theta_k$,
$\phi_{\sum_{k=1}^S \Theta_k}(t)$, satisfies
\begin{equation}
\phi_{\sum_{k=1}^S \Theta_k}(t)=\prod_{k=1}^{S}\phi_{\Theta_k}(t).
\end{equation}
In particular, for the case of two summands, $S=2$, $\phi_{\Theta_1+\Theta_2}(t)=\phi_{\Theta_1}(t)\phi_{\Theta_2}(t)$ and for the NNTS case,
\begin{eqnarray*}
\phi_{\Theta_1+\Theta_2}(t)& = & \left(I(t=0) + \sum_{k=-M_1}^{-1} \left(\sum_{j=0}^{M_1-k}c_j^{(1)}\bar{c}_{j+k}^{(1)}\right) I(t=k) + \sum_{k=1}^{M_1}
\left(\sum_{j=0}^{M_1-k}c_{j+k}^{(1)}\bar{c}_{j}^{(1)}\right) I(t=k) \right) \\
 & \times & \left(I(t=0) + \sum_{k=-M_2}^{-1} \left(\sum_{j=0}^{M_2-k}c_j^{(2)}\bar{c}_{j+k}^{(2)}\right) I(t=k) + \sum_{k=1}^{M_2}
 \left(\sum_{j=0}^{M_2-k}c_{j+k}^{(2)}\bar{c}_{j}^{(2)}\right) I(t=k) \right) \\
\end{eqnarray*}
Finally, obtaining
\begin{eqnarray*}
\phi_{\Theta_1+\Theta_2}(t)& = & I(t=0) + \sum_{k=-\min\{M_1,M2\}}^{-1} \left(\sum_{j=0}^{\min\{M_1,M-2\}-k}c_j^{(1)}\bar{c}_{j+k}^{(1)}c_j^{(2)}\bar{c}_{j+k}^{(2)}\right)
I(t=k) + \\
 & & \sum_{k=1}^{\min\{M_1,M_2\}} \left(\sum_{j=0}^{\min\{M_1,M_2\}-k}c_{j+k}^{(1)}\bar{c}_{j}^{(1)}c_{j+k}^{(2)}\bar{c}_{j}^{(2)}\right) I(t=k). \\
\end{eqnarray*}
Extending this result to the case of $S$ summands, the characteristic function of the sum $\sum_{k=1}^S \Theta_k$ of the independent NNTS circular random variables is given by
\begin{eqnarray*}
\phi_{\sum_{k=1}^S \Theta_k}(t) & = & I(t=0) + \\
 & & \sum_{k=-\min\{M_1,M2, \ldots, M_S\}}^{-1} \left(\sum_{j=0}^{\min\{M_1,M-2, \ldots, M_S\}-k}\prod_{s=1}^{S}c_j^{(s)}\bar{c}_{j+k}^{(s)}\right) I(t=k) + \\
 & & \sum_{k=1}^{\min\{M_1,M_2, \ldots, M_S\}} \left(\sum_{j=0}^{\min\{M_1,M_2, \ldots, M_S\}-k}\prod_{s=1}^{S}c_{j+k}^{(s)}\bar{c}_{j}^{(s)}\right) I(t=k). \\
\end{eqnarray*}
Thus, the NNTS family of circular distributions is closed under summation; that is, the sum of independent NNTS circular random variables is an NNTS circular random variable
with parameter $M_{sum}=\min\{M_1, M_2, \ldots, M_S\}$ and parameter $\underline{c}^{sum}$, which is a function of the vectors of parameters $\underline{c}^{(s)}$, $s=1,
\ldots, S$.

To obtain the vector of parameters $\underline{c}^{sum}$, the following system of $M_{sum}$ equations involving the real parameter $c_0^{sum}$ is considered.
\begin{equation}
\prod_{s=1}^{S}c_{k}^{(s)}\bar{c}_{0}^{(s)} = c_0^{sum}c_{k}^{sum}
\label{csum}
\end{equation}
for $k=1, \ldots, M_{sum}$ and the norm equation
\begin{equation}
(c_{0}^{sum})^2 + \sum_{k=1}^{M_{sum}} ||c_{k}^{sum} ||^2 = 1.
\label{norm}
\end{equation}

By considering the real and imaginary parts in Equations (\ref{csum}) and (\ref{norm}), this system of $2M_{sum}+1$ nonlinear real equations can be solved numerically. In
particular, one can use the $R$ package $rootSolve$ (Soetaert, 2009) considering the vector with one in its first entry and zeroes in the other entries
that correspond to the uniform distribution case as initial values.

For the sum of more than two NNTS circular random variables, the result for two random variables can be applied recursively.

\subsection{Case $M=$1}

If both summands are NNTS random variables with $M=1$, then the density function of their sum can be obtained analytically as follows:
By considering the squared norm in the equations defined in Equation (\ref{csum}), we obtain
\begin{equation}
|| \prod_{s=1}^{S}c_{k}^{(s)}\bar{c}_{0}^{(s)} ||^2= (c_0^{sum})^2||c_{k}^{sum}||^2.
\label{csum2}
\end{equation}
Substituting these equations into Equation (\ref{norm}), the following equation for $c_{0}^{sum}$ is obtained:
\begin{equation}
(c_{0}^{sum})^2 + \sum_{k=1}^{M_{sum}} \frac{|| \prod_{s=1}^{S}c_{k}^{(s)}\bar{c}_{0}^{(s)} ||^2}{(c_{0}^{sum})^2} = 1
\end{equation}
which is equivalent to the following biquadratic equation on $c_{0}^{sum}$:
\begin{equation}
(c_{0}^{sum})^4 - (c_{0}^{sum})^2 + \sum_{k=1}^{M_{sum}} || \prod_{s=1}^{S}c_{k}^{(s)}\bar{c}_{0}^{(s)} ||^2 = 0
\end{equation}
with the largest positive solution given by
\begin{equation}
c_0^{sum} = \sqrt{\frac{1+\sqrt{1-4\sum_{k=1}^{M_{sum}} || \prod_{s=1}^{S}c_{k}^{(s)}\bar{c}_{0}^{(s)} ||^2}}{2}}.
\end{equation}
Once the value of $c_0^{sum}$ is determined, the values of $c_1^{sum}, \ldots, c_{M_{sum}}^{sum}$ can be obtained by using the system of equations in Equation (\ref{csum}).

Thus, for the sum of two NNTS circular random variables with $M=$1, the $c$ parameters are given by
\begin{equation}
c_0^{sum}=\sqrt{\frac{1+\sqrt{1-4 || c_{1}^{(1)}\bar{c}_{0}^{(1)}c_{1}^{(2)}\bar{c}_{0}^{(2)} ||^2}}{2}}
\end{equation}
and
\begin{equation}
c_1^{sum}=\frac{c_{0}^{(1)}c_{1}^{(2)}c_{0}^{(2)}}{c_0^{sum}}.
\end{equation}

\subsection{Numerical Examples}

In the case of the sum of two NNTS circular random variables with different values of $M$, Figures \ref{graphsumMigual} and \ref{graphsumMdif} show the density functions of the two random variables and their sum. In addition, the horizontal line corresponding to the circular uniform density ($M=0$) was included to appreciate the convergence of the sum to the circular uniform density. The plots on the right of Figures \ref{graphsumMigual} and \ref{graphsumMdif} include the histograms of 1000 realisations from the sum of the two univariate NNTS densities; considering realisations from each of the summands and then their sum (modulus 2$\pi$), the NNTS density of the sum is superimposed on the histograms.

\renewcommand{\tablename}{Figure}
\begin{table}[t]
\centering
\begin{tabular}{cc}
\rotatebox{90}{\includegraphics[scale=.4,angle=270,origin=c]{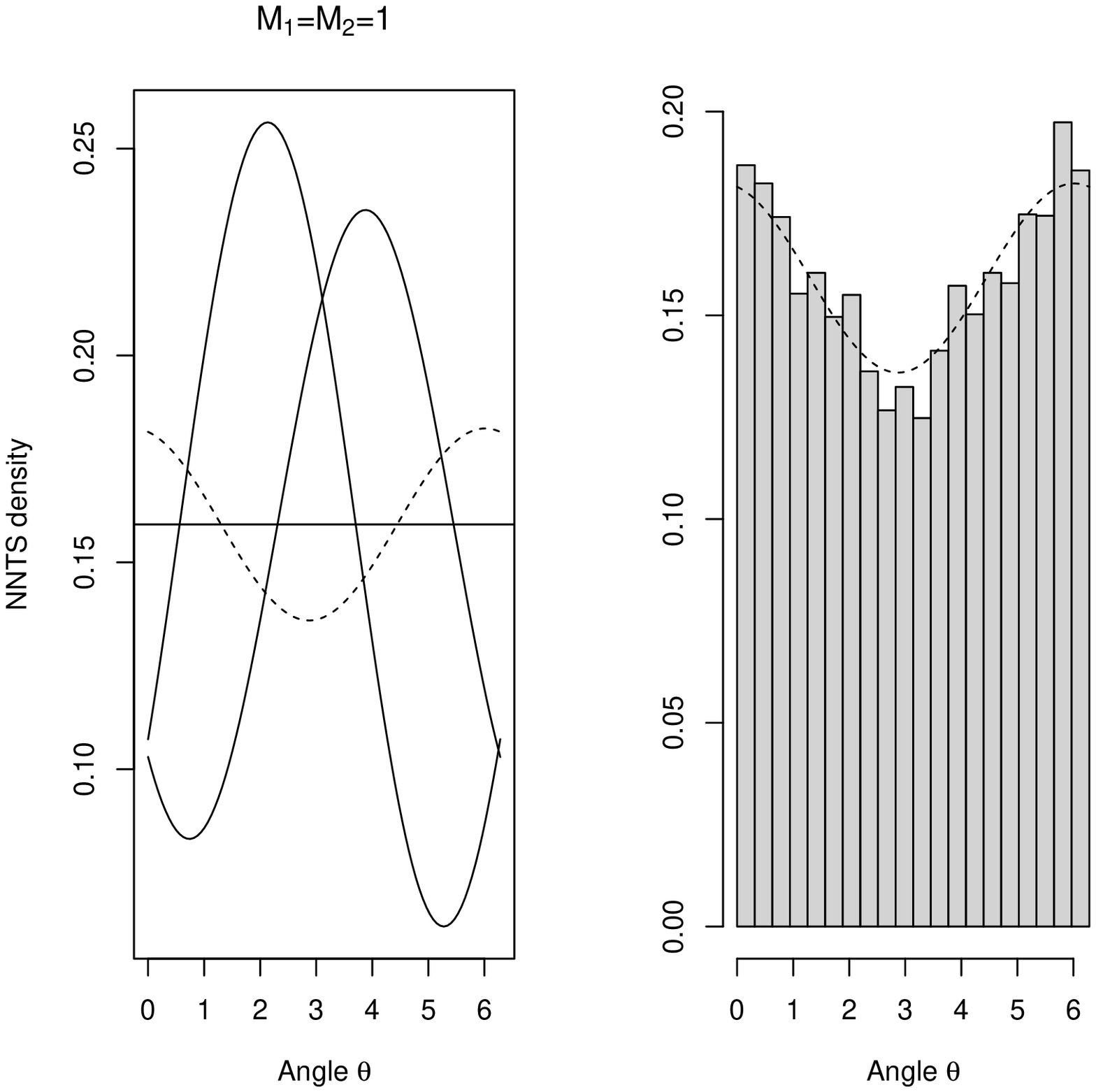}} & \rotatebox{90}{\includegraphics[scale=.4,angle=270,origin=c]{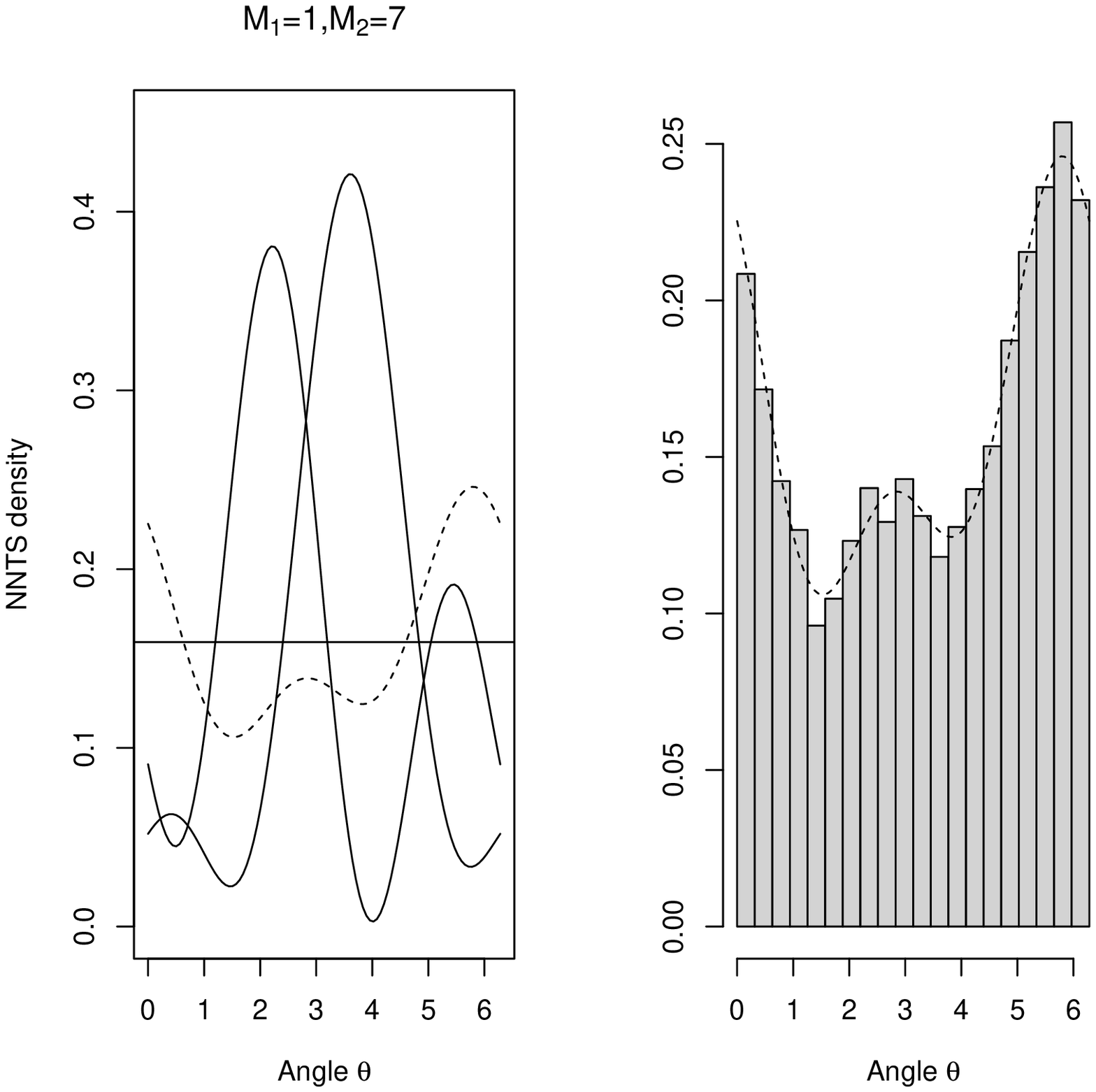}} \\
\rotatebox{90}{\includegraphics[scale=.4,angle=270,origin=c]{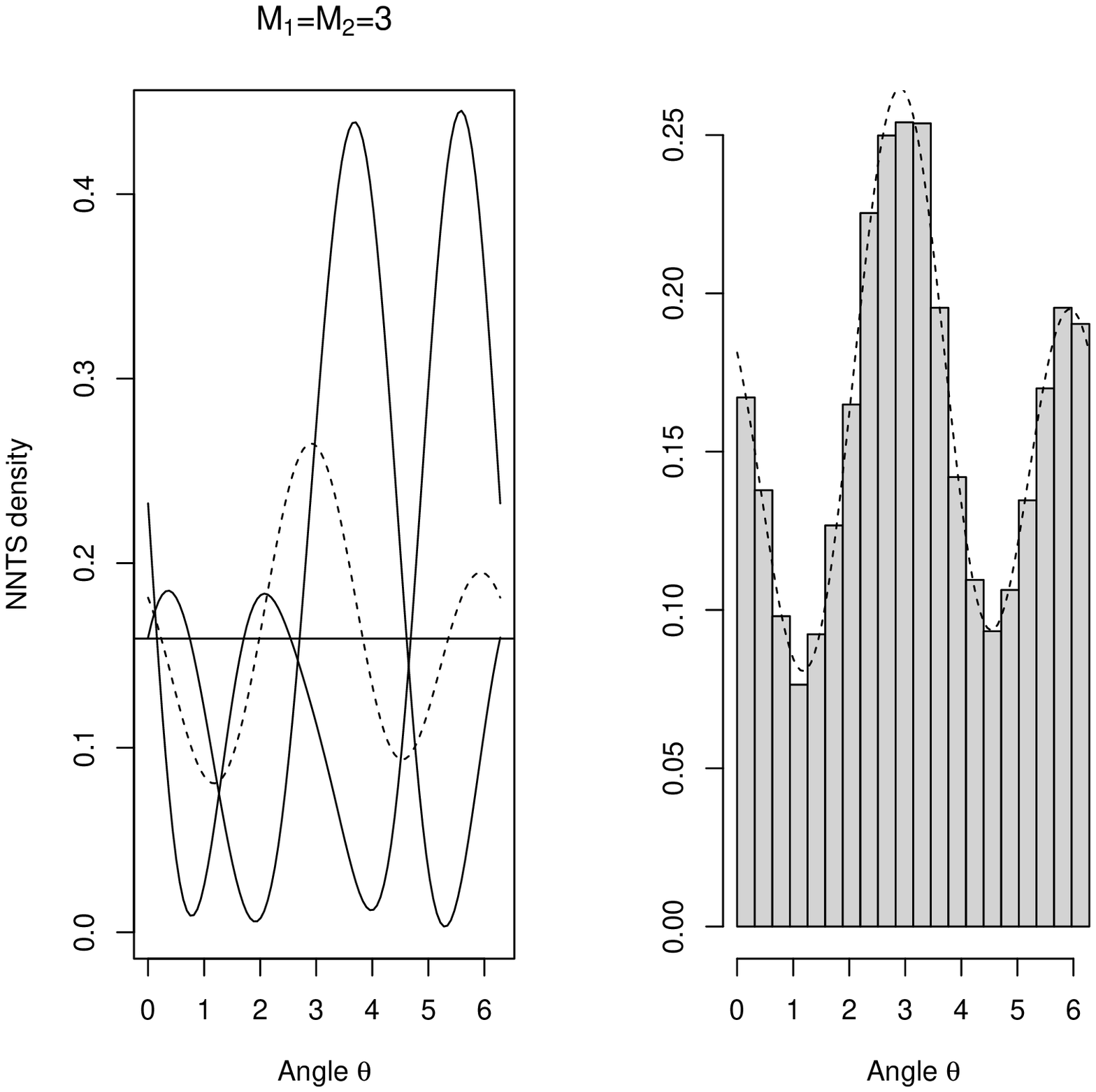}} & \rotatebox{90}{\includegraphics[scale=.4,angle=270,origin=c]{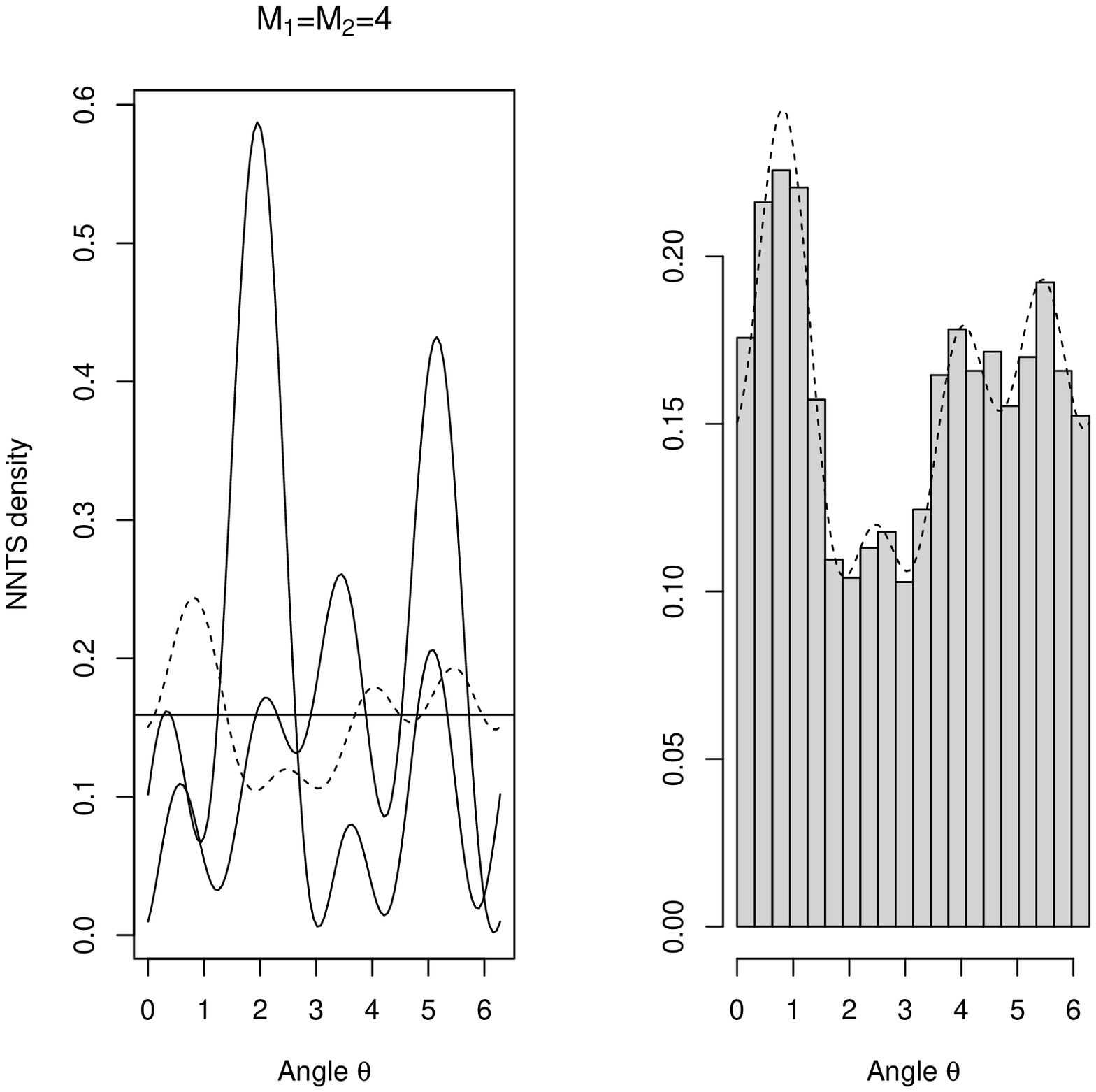}} \\
\end{tabular}
\caption{Examples of plots of the density functions of the sum of two NNTS elements with the same value of $M$ for $M$=1,2,3, and 4. The left plots show the plots of the NNTS densities of the variables in the sum and the NNTS density of the resulting sum. The right plots show the histograms of 1000 realisations from the resulting NNTS model of the sum superimposed with the NNTS density of the sum.
\label{graphsumMigual}}
\end{table}

\renewcommand{\tablename}{Figure}
\begin{table}[t]
\centering
\begin{tabular}{cc}
\rotatebox{90}{\includegraphics[scale=.4,angle=270,origin=c]{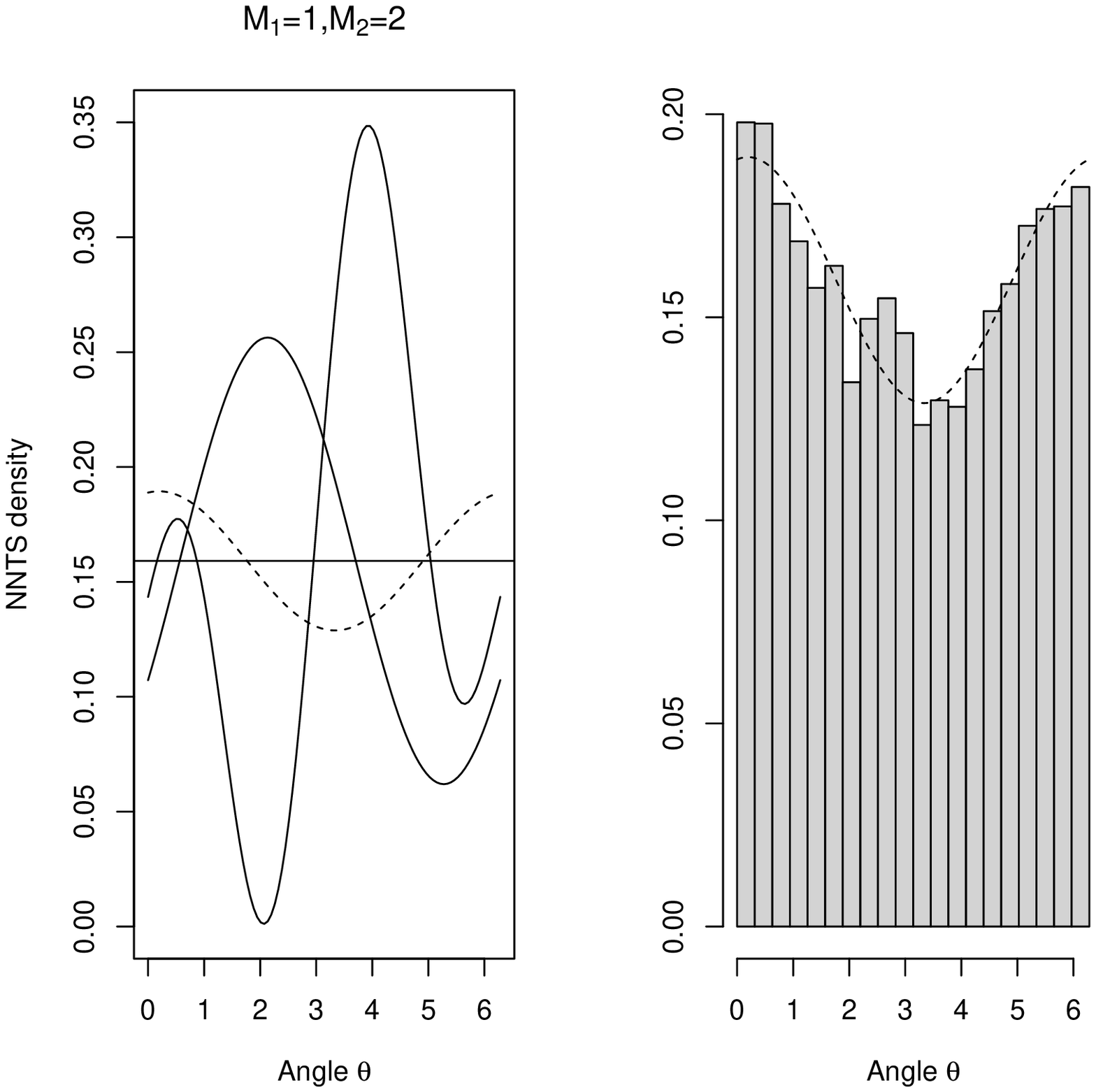}} & \rotatebox{90}{\includegraphics[scale=.4,angle=270,origin=c]{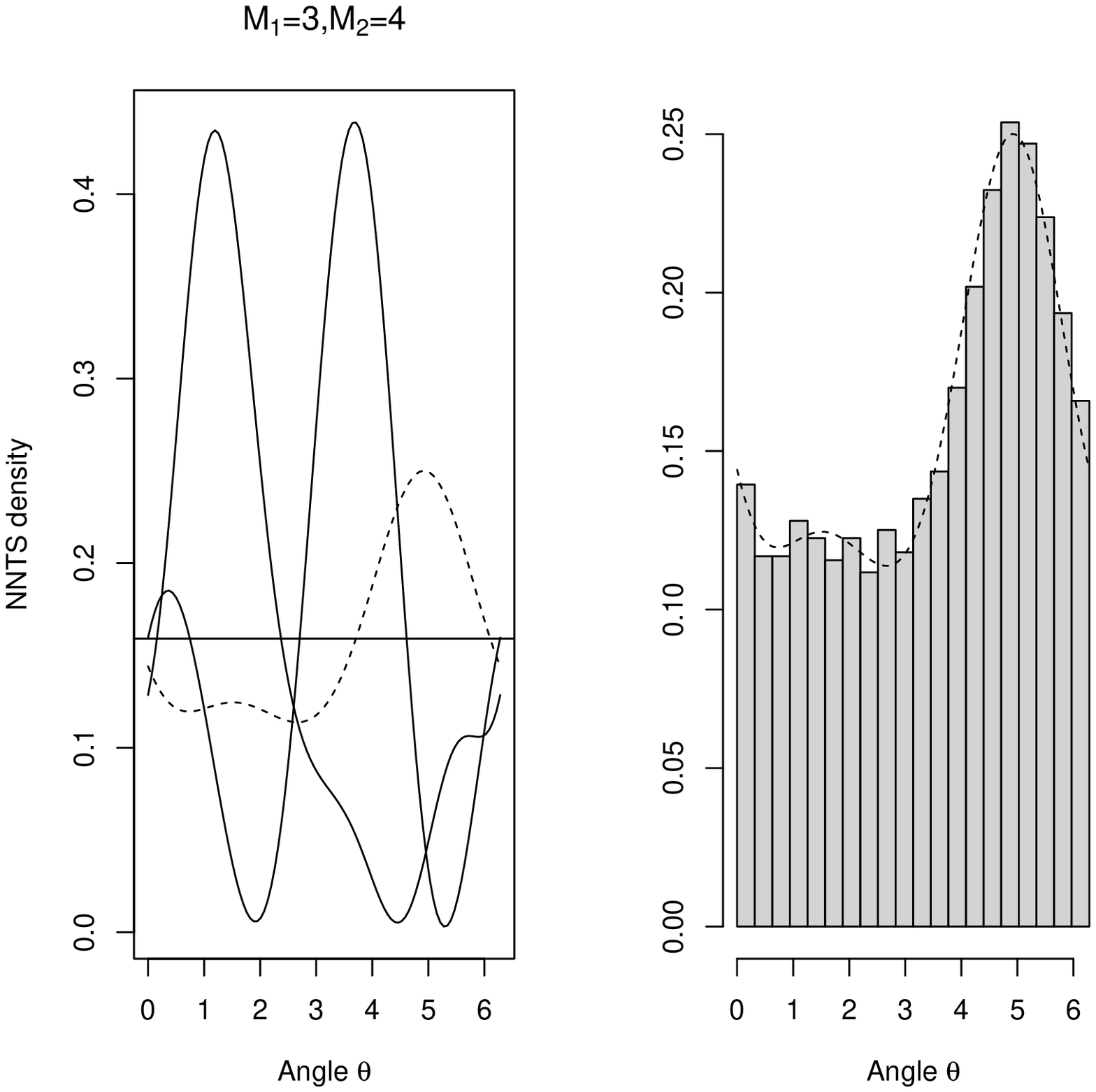}} \\
\rotatebox{90}{\includegraphics[scale=.4,angle=270,origin=c]{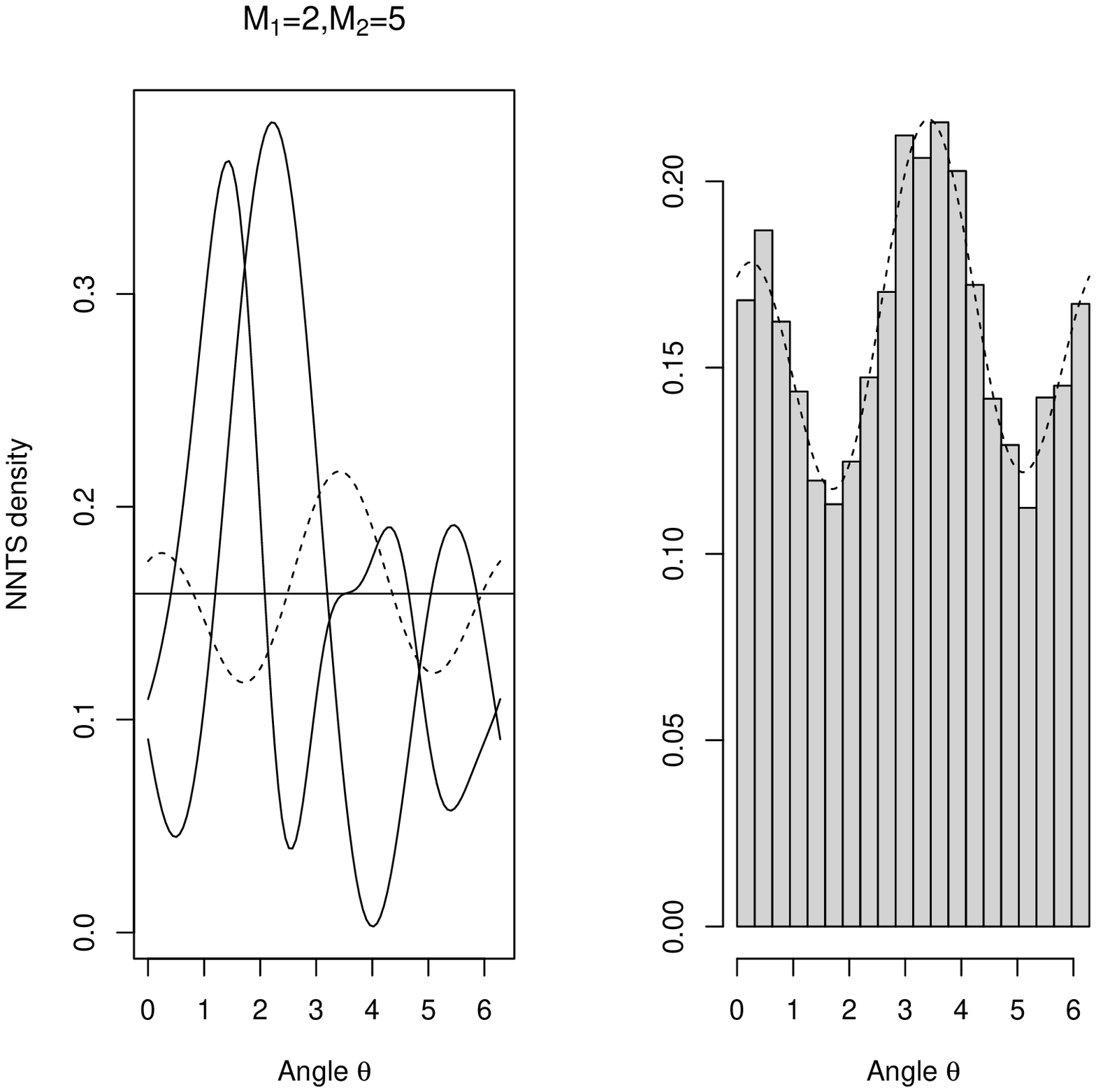}} & \rotatebox{90}{\includegraphics[scale=.4,angle=270,origin=c]{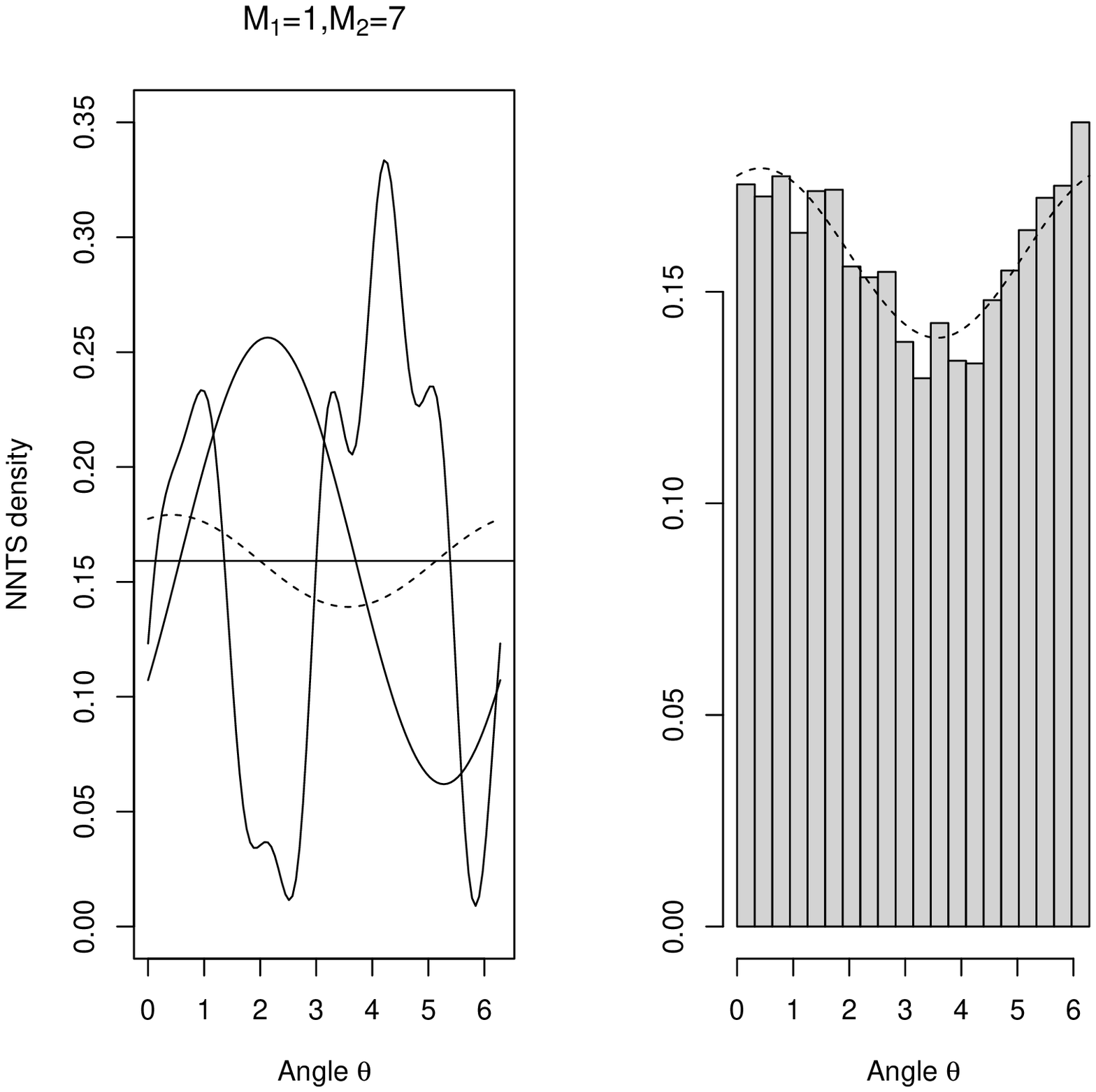}} \\
\end{tabular}
\caption{Examples of plots of the density functions of the sum of two NNTS elements with different values of $M$, $M_1$ and $M_2$, for the combinations $(M_1,M_2)$=(1,2), (3,4), (2,5), and (1,7). The right plots show the histograms of 1000 realisations from the resulting NNTS model of the sum superimposed with the NNTS density of the sum.
\label{graphsumMdif}}
\end{table}

Figure \ref{graphsumiid} presents, for a simulated case with $M=5$, the plots of the NNTS densities for the case of independent and identically distributed random variables, in which we add recursively to obtain the density function of the sum of 2, 3, 4, 5, and 6 random variables. From Figure \ref{graphsumiid}, it is clear how the convergence to the circular uniform distribution occurs very fast, with the sum of three or more random variables appearing almost circularly uniformly distributed.

\renewcommand{\tablename}{Figure}
\begin{table}[h]
\centering
\begin{tabular}{c}
\includegraphics[scale=.8, bb=0 0 504 504]{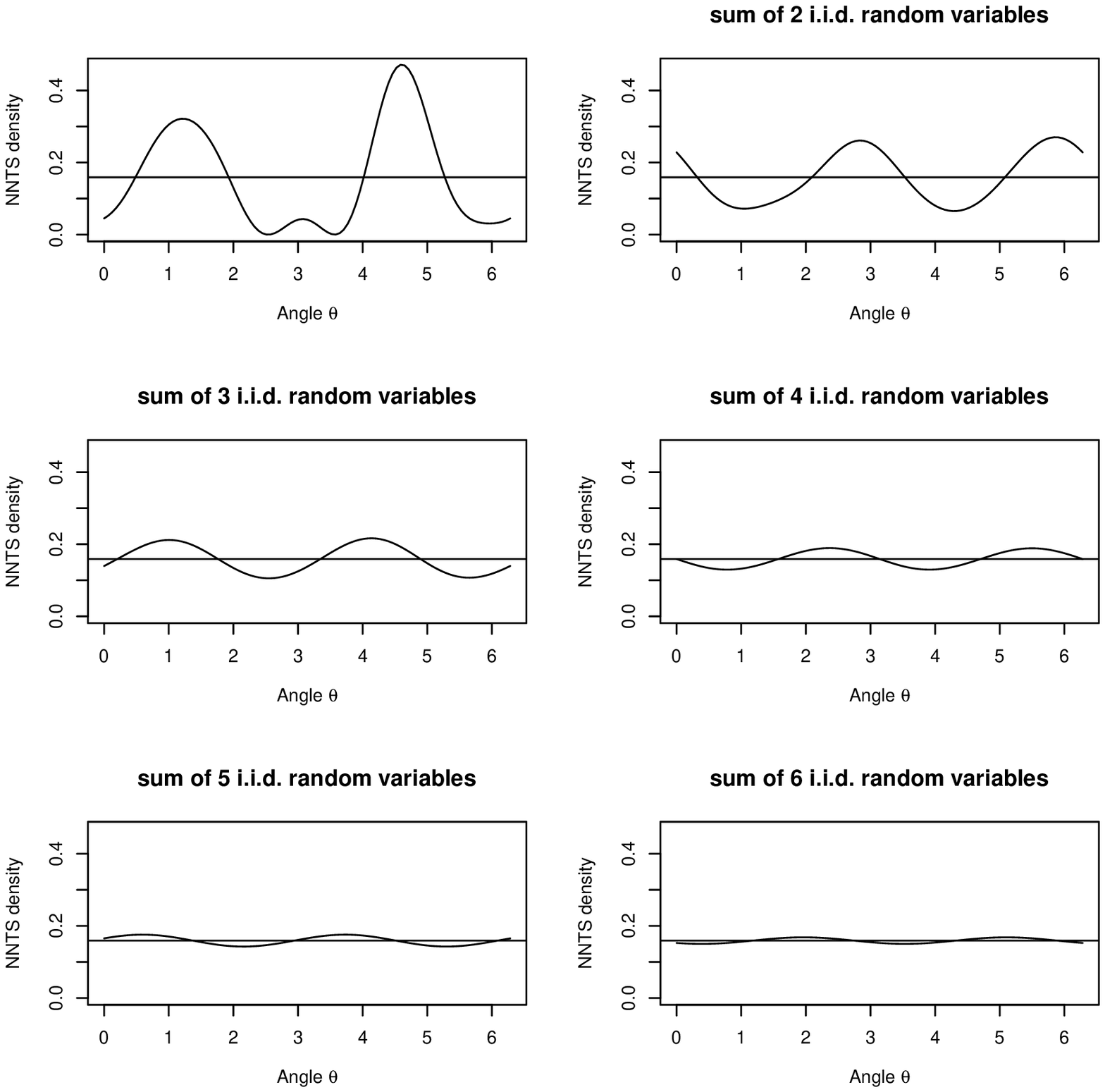} \\
\end{tabular}
\caption{Examples of plots of densities of sums of i.i.d. NNTS random variables with $M$=5. The first plot shows the NNTS density with $M$=5.
\label{graphsumiid}}
\end{table}

\section{Two Circular Uniformity Tests with NNTS as Alternative Hypotheses}

Many tests for circular uniformity have been reported in the literature. Among the most used in practice (see Fisher, 1993; Mardia and Jupp, 2000; and Upton and Fingleton, 1989), one finds the Rayleigh test against a unimodal alternative, Kuiper's test, Watson's test, and the range test among many others.
As noted by Fisher (1993, p. 65), circular uniformity tests depend on the
specification of the model in the alternative hypothesis, and one wants to have an alternative model that has a different number of modes to detect
any departure from uniformity. Given that the family of NNTS circular distributions is nested, that is, all models with $M=M^*$ are particular cases of NNTS circular
distributions with $M=M^{**}$ with $M^{**}>M^*$, NNTS circular distributions are suitable models for detecting any departure from uniformity for a sufficiently large
sample size. Various studies have been conducted on the low power of many circular uniformity tests. For example, Landler et al. (2018) compared the power of the Rayleigh
test, Watson's test, Kuyper's test, Rao's spacing test, Bogdan test, and
Hermans-Rasson test. Their main conclusions are that the Rayleigh test is preferred
for unimodal departures from circular uniformity and that for multimodal departures from circular uniformity, the Hermans-Rasson test is recommended when considering mixtures
of von Mises distributions as alternative models and eight different sample sizes (10, 15, 20, 25, 30, 40, 80, and 100). In the case of symmetric multimodality, the
transformation of the data to a unimodal distribution and application of the Rayleigh test is recommended. Later, Landler et al. (2019) compared the power of the
Rayleigh test, the Hermans-Rasson original test, a modification of the Hermans-Rasson test, and the Pycke test when considered as alternative model mixtures of von Mises
distributions with modes equally distributed in the interval $[0,2\pi)$ with different proportions assigned to the different elements of the mixture and a sample size
of 60. The final recommendations are to use the Rayleigh tests for unimodal departures from circular uniformity, the original Hermans-Rasson test for alternative
distributions with at least two modes, and when the sample size is large and one considers at least two modes in the alternative distribution, the recommendation is to use the Pycke test
instead. In addition, they point to the difficulty of testing for circular uniformity when the number of modes is greater than two and the alternative distribution is unknown and recommend substantially increasing the sample size and running the Pycke test with the constraint that the observed angles are supposed to show a high
concentration around the modes. Given these results, we compared the power of the two proposed NNTS tests against four different tests: the Rayleigh test, the
original and modified Hermans-Rasson tests, and the Pycke test. For a random sample of angles $\theta_1, \theta_2, \ldots, \theta_n$, the test statistic, as presented in
Landler et al. (2019), of the original Hermans-Rasson test is
\begin{equation}
T_{HRo}=\frac{n}{\pi} - \left(\frac{1}{2n}\right)\sum_{i=1}^{n}\sum_{j=1}^{n}|\sin(\theta_1-\theta_j)|,
\end{equation}
the modified Hermans-Rasson test is
\begin{equation}
T_{HRm}=\left(\frac{1}{n}\right)\sum_{i=1}^{n}\sum_{j=1}^{n}\left(||\theta_i-\theta_j|-\pi| - \frac{\pi}{2} - 2.895(|\sin(\theta_i - \theta_j)|-\frac{2}{\pi})\right)
\end{equation}
while the Pycke test is
\begin{equation}
T_{P}=\left(\frac{1}{n}\right)\sum_{i=1}^{n}\sum_{j=1}^{n} \left(\frac{2(\cos(\theta_i-\theta_j)-\sqrt{0.5})}{1.5 - (2\sqrt{0.5}\cos(\theta_i-\theta_j))}  \right).
\end{equation}
The Hermans-Rasson tests belong to the family of circular uniformity tests of Beran (1969), also known as Sobolev's tests, in which the mean resultant length of the observations is p-fold wrapped on the unit circle, which is equivalent to considering the powers of the unit complex vectors with arguments given by the observed angles and calculating their mean resultant lengths, thereby obtaining weighted sums of Rayleigh statistics for different powers (see Mardia and Jupp, 2000).
Like this study, Pycke (2010) considered multimodal alternative distributions that correspond to Fourier transformations that are equivalent to NNTS densities but Pycke did not consider the constraints in the parameter space to obtain a valid density function that is positive and integrates to one. Pycke (2010) found that the distribution of his test statistic is of a nonstandard form and corresponds to a weighted sum of chi-square distributions in which the weights are unknown complex functions of the observations.
Given the convergence of the distribution of the sum of circular random variables to the circular uniform distribution and, for the case of NNTS circular random variables, it is possible to investigate and
compare the properties of the proposed NNTS test for cases where the parameter vector is close to the value specified to the null hypothesis.
A circular uniformity test with NNTS distributions as alternative hypotheses exploits the flexibility of NNTS densities, which can model
very different patterns for the alternative distribution in terms of the number of modes and asymmetry. For the NNTS test, the null and alternative hypotheses were specified as follows:
\begin{equation}
H_0: M = 0 \mbox{ vs. } H_a:M=M^* > 0.
\end{equation}
or, equivalently,
\begin{equation}
H_0: f_{\Theta}(\theta)=\frac{1}{2\pi} \mbox{ vs. } H_a: f_{\Theta}(\theta) = \frac{1}{2\pi}\left|\left|\sum_{k=0}^{M^*} c_k  e^{ik\theta}\right|\right|^2 =
\frac{1}{2\pi}\sum_{k=0}^{M^*}\sum_{m=0}^{M^*} c_k\bar{c}_m e^{i(k-m)\theta}.
\end{equation}
In terms of the $\underline{c}$ parameter vector of an NNTS density with a fixed value of $M=M^*$, the null and alternative hypotheses are as follows:
\begin{equation}
H_0: \underline{c}=(1,0,0, \ldots, 0)^{\top} \mbox{ vs. } H_a: \underline{c} \ne (1,0,0, \ldots, 0)^{\top}.
\end{equation}
This hypothesis test is nonregular because the null hypothesis specifies the parameter vector on the boundary of the parameter space, and
the maximum likelihood asymptotic results under regularity conditions do not apply. In particular, the likelihood ratio test statistic does not converge in distribution to
a chi-squared distribution, and common bootstrap procedures are not applicable. Because the null hypothesis corresponds to the circular uniform distribution,
the critical values of the NNTS test are obtained by simulating samples from the circular uniform distribution and, for each sample, fitting the NNTS
model specified under the alternative hypothesis by maximum likelihood to calculate the value of the test statistic. For the first NNTS test for circular uniformity ($NNTS1$), we considered test statistic the standardised maximum
likelihood estimator, $\hat{\underline{c}}$, of the vector of parameters $\underline{c}$ defined as:
\begin{equation}
T_{NNTS1}=(\hat{\underline{c}}-\underline{c})^{H}j(\hat{\underline{c}})(\hat{\underline{c}}-\underline{c})
\end{equation}
where $\underline{c}^{H}$ is the Hermitian (conjugate and transpose) of the vector $\underline{c}$ and $j(\hat{\underline{c}})$ is the observed information that is proportional to the Hessian matrix that includes the second derivatives of the log-likelihood function,
and for the NNTS density, it is equal to the projection matrix (see Fern\'{a}ndez-Dur\'{a}n and Gregorio-Dom\'{i}nguez, 2010)
\begin{equation}
j(\hat{\underline{c}})=n(\mathbb{I}-\hat{\underline{c}}\hat{\underline{c}}^{H})
\end{equation}
where $n$ is the sample size and $\mathbb{I}$ is the $(M+1)\times(M+1)$ identity matrix. Because $j(\hat{\underline{c}})$ is a projection matrix that is not an identity
matrix, it is not invertible, making this a nonregular maximum likelihood estimation problem. Then,
\begin{equation}
T_{NNTS1}=
n(\hat{\underline{c}}-\underline{c}_0)^{H}(\mathbb{I}-\hat{\underline{c}}\hat{\underline{c}}^{H})(\hat{\underline{c}}-\underline{c}_0) =
n(\hat{\underline{c}}-\underline{c}_0)^{H}(\hat{\underline{c}}-\underline{c}_0)-
n(\hat{\underline{c}}-\underline{c}_0)^{H}(\hat{\underline{c}}\hat{\underline{c}}^{H})(\hat{\underline{c}}-\underline{c}_0)
\end{equation}
By partitioning the maximum likelihood estimator as $\hat{\underline{c}}^H=(\hat{c}_0,\hat{c}_1,\ldots,\hat{c}_M)^H=(\hat{c}_0,\hat{c}_{1:M})^H$ with
$\hat{c}_{1:M}^H=(\hat{c}_1, \ldots, \hat{c}_M)^H$ and considering that $\sum_{k=0}^M ||\hat{c}_k||^2=1$, one obtains $(\hat{\underline{c}}-\underline{c}_0)^{H}(\hat{\underline{c}}-\underline{c}_0)=(\hat{c}_0-1)^2 + 1 - \hat{c}_0^2$ and $(\hat{\underline{c}}-\underline{c}_0)^{H}(\hat{\underline{c}}\hat{\underline{c}}^{H})(\hat{\underline{c}}-\underline{c}_0)=(1-\hat{c}_0)^2$. Then,
\begin{equation}
T_{NNTS1}(\hat{\underline{c}})=n(1-\hat{c}_0^2).
\end{equation}
$T_{NNTS1}$ depends only on the first component of the maximum likelihood vector and, intuitively, because the sum of the norms of the
components of the parameter vector, $\underline{c}$, should be equal to one, $T_{NNTS1}$ measures (scaled by the sample size) how far is $\hat{c}_0^2$
of being equal to one that corresponds to the circular uniform distribution case.

Table \ref{criticalvalues} lists the critical values for $T_{NNTS1}$ obtained by the simulation for significance levels of 10\%, 5\%, and 1\% for
different sample sizes. We used a total of 10000 simulated samples to obtain critical values. Given the recommendations in Cuddington and Navidi (2011) for the number of simulated samples to produce critical values, the critical values in Tables \ref{criticalvalues} and \ref{criticalvaluesloglik} are reported with a precision of 0.1.

The second maximum likelihood NNTS test for circular uniformity is based on the generalised likelihood ratio statistic defined as
\begin{equation}
T_{NNTS2}=-2\ln\left(\frac{\hat{L}_{H_0}}{\hat{L}_{H_a}}\right)=-2\ln\left(\frac{\hat{L}_{M=0}}{\hat{L}_{M=M^*}}\right)=2\ln(\hat{L}_{M=M^*})+2n\ln(2\pi)
\end{equation}
where $\hat{L}_{M=M^*}$ is the maximised likelihood under the alternative hypothesis $H_a: M=M^*$, which corresponds to the maximised likelihood of the NNTS model with
$M=M^*>0$. Again, because the maximum likelihood of the NNTS model does not satisfy the regularity conditions under the null hypothesis of uniformity, the critical values
are obtained by simulation and are included in Table \ref{criticalvaluesloglik} for various values of $M$ (1, 2, ..., 7), significance levels $\alpha$ (10\%, 5\%, and 1\%), and various sample sizes. Again, given the nonregular maximum likelihood estimation for NNTS models under the null hypothesis ($M=0$), the statistic
$T_{NNTS2}$ does not converge to a chi-squared distribution for large sample sizes, and commonly used bootstrap procedures are not applicable. Table \ref{criticalvaluesloglik} contains a larger number of sample sizes than Table \ref{criticalvalues} since, as shown later in the paper, the NNTS2 has more power than the NNTS1 test and is recommended for use in practice.

Following MacKinnon (1991 and 2010), Table \ref{criticalvaluesregressionmodels} includes the fitted regression models to interpolate the critical values for any sample size with a precision of 0.1 (one decimal place). In this case, the regression models for the critical values considered as explanatory variables the reciprocal of the sample size and the NNTS parameter $M$ and their interaction and the reciprocal of the squared sample size. The interaction between the squared sample size and $M$ was not significant for all the considered models. Initially, a single regression model for all the values of $M$ was considered, but for the cases $M=$1 and $M=$2, it did not present a good fit. Then, two separate regression models were fitted for the cases $M=$1 and $M=$2, in which only the sample size and $M$ had significant coefficients. For the other considered values of $M$, 3 to 7, a common regression model was sufficient. As shown in Table \ref{criticalvaluesregressionmodels}, the fitted regression models had a good fit since their coefficients of determination are very high and their maximum absolute and relative errors are quite small. The relative errors are less than 2.1\% for the model with $M=$1 and less than 1.3\% for the other models ($M \ge 2$). Given these results, the critical values for all sample sizes can be interpolated for any sample size by using the fitted regression models. Given the observed precision of the regression models, in the case of an observed NNTS2 statistic, $T_{NNTS2}$, with a value that differs from the interpolated critical value by less than 0.1, the test can be considered inconclusive. Table \ref{criticalvaluesregressionmodels} also includes the sample sizes at which the interpolated critical values by regression reach the asymptotic critical values observed in the simulations in Table \ref{criticalvaluesloglik}. From these identified sample sizes, the asymptotic values obtained in the simulation are used in the implementation of the test. These asymptotic critical values were determined in the simulations by identifying many consecutive sample sizes at which the critical values obtained by simulation did not change. For the fitting of the regression models, we considered only the first two consecutive sample sizes at which the critical values did not change. From the simulations and by considering the critical values as a decreasing function of the sample size, the minimum sample size to apply the NNTS2 test for $M \ge$3 was found to be $10(M+1)$ which implies that we have at least 5 observations for each of the 2$M$ NNTS parameters to be estimated. For cases $M=$1 and $M=$2, we found that the required minimum sample sizes are 15 and 25, respectively.

\section{Power and Size Comparisons}

We compared the Rayleigh (RT), modified Hermans-Rasson (HRmT), Pycke (PT), and NNTS ($NNTS1$ and $NNTS2$) tests in terms of their power and size by simulating samples from
the null circular uniform distribution and the alternative NNTS distribution for sample sizes (SS) of 25, 50, 100, 200, and 500. We compared the power of the tests for
significance levels $\alpha$ of 10\%, 5\%, and 1\%. The $R$ package $circular$ (Agostinelli and Lund, 2017) was used to calculate the test statistic of the
Rayleigh test. The Hermans-Rasson and Pycke tests were performed by using the $R$ package $CircMLE$ (Fitak and Johnsen, 2017; and Landler et al., 2019). Finally, for the
$NNTS1$ and $NNTS2$ the $R$ package $CircNNTSR$ (Fern\'{a}ndez-Dur\'{a}n and Gregorio-Dom\'{i}nguez, 2016) was used. To speed up the calculation of the NNTS tests, the
computations were implemented in parallel by using the $R$ package $parallel$ (R Core Team, 2021) in an 8 core CPU at a speed of 3 GHz.

Figure \ref{graphpowerexamples} shows plots of the two NNTS alternative models with $M=3$ and $M=6$. For each of the two NNTS alternative models, we considered various values of the parameter $c_0$ to obtain alternative models that are close to the null circular uniform distribution. As shown in Figure
\ref{graphpowerexamples}, by increasing the value of parameter $c_0$, we obtain distributions that are closer to the circular uniform distribution. In terms of size, when
simulating samples from the null circular uniform distribution and applying the tests, all the considered tests obtained an adequate observed frequency of rejection of the null
hypothesis that was practically identical to the significance level. We used 1000 simulated samples from the null and alternative models in our simulations, and the frequencies corresponding to the observed power are reported in rounded percentages ranging from 0\% to 100\%.

Tables \ref{powervalues10}, \ref{powervalues05}, and \ref{powervalues01} contain the results for the power of the tests by using the observed frequencies, in percentage, for
rejecting the null hypothesis of circular uniformity when simulating random samples from the alternative model with $M=3$ with eight different values of
parameter
$c_0$. The considered eight cases of the $c_0$ parameter range from 0.59 to 0.9959, with $c_0$ values near one representing densities closer to the circular uniform
distribution, as shown in the left plot in Figure \ref{graphpowerexamples}. Basically, in all cases and sample sizes where the power takes an acceptable value, the
Hermans-Rasson (HRmT) test presented lower power than that for the Pycke (PT) test, and we then compared the NNTS ($NNTS1$ and $NNTS2$) tests against the Pycke (PT)
test. For the sample size of 25, the Pycke test has the largest power, although it is below 0.6. For cases 1 to 5 and sample sizes of 25 and 50, the $NNTS2$ with $M=3$ had the largest power, followed by the $NNTS2$ test with $M=4$, which was followed by the
Pycke test. In many cases and sample sizes, the $NNTS2$ test with $M=5$ has a very similar power to the Pycke test. This implies that when applying the generalised
likelihood ratio $NNTS2$ test, there is some flexibility in the selection of the $M$ value to be used in the test; in case of doubt between $M=M^*$ and $M=M^*+1$, it is
recommended using $M=M^*+1$ to avoid a situation in which a smaller $M$ is used and the power decreases considerably, as shown for the power values for the $NNTS2$ test with
$M=2$ or $M=1$. For the largest sample sizes of 200 and 500 in Tables \ref{powervalues05} and \ref{powervalues01}, $NNTS1$ and $NNTS2$ present similar power values,
showing that the two tests are equivalent for large sample sizes and significance levels of 5\% and 1\%, respectively. This convergence was achieved earlier for sample sizes of 100, 200,
and 500 for a significance level of 10\%, as shown in Table \ref{powervalues10}.

For cases 6, 7 and 8 and sample sizes of 25, 50, and 100, none of the tests showed acceptable power, implying that a larger sample size is required to detect small deviations from circular uniformity. For example, for case 6 with $c_0=0.9899$, one obtains acceptable power for the $NNTS2$ test with $M=3,4, \mbox{ or } 5$ only for a sample
size equal to 500.

Table \ref{powervaluesM6} presents a comparison of the generalised likelihood ratio $NNTS2$ test with $M=6$ and $M=7$ and the Pycke test for simulated data from the NNTS
alternative model with $M=6$ and six cases with values of the parameter $c_0$ from 0.5072892 to 0.9999601 presented in the right plot in Figure
\ref{graphpowerexamples}. For cases 4, 5 and 6, it is clear from the low power of the tests that sample sizes larger than 500 are required to detect very small
deviations from circular uniformity implied by the $c_0$ values that are very close to one. For cases 1, 2 and 3, the $NNTS2$ tests with $M=6$ are the ones that almost in
all cases, present the largest power followed by the $NNTS2$ test with $M=7$, and this test is followed by the Pycke test. The difference between the powers of the $NNTS2$
test and the Pycke test can be large, as shown in case 3. Again, the use of the $NNTS2$ test is recommended, and the value of $M$ can be larger than the true value, and still
one obtains a larger power than that for the Pycke test.

\renewcommand{\tablename}{Figure}
\begin{table}[t]
\centering
\begin{tabular}{c}
%\rotatebox[origin=c]{-90}{
\includegraphics[scale=.75, bb=54 144 558 648]{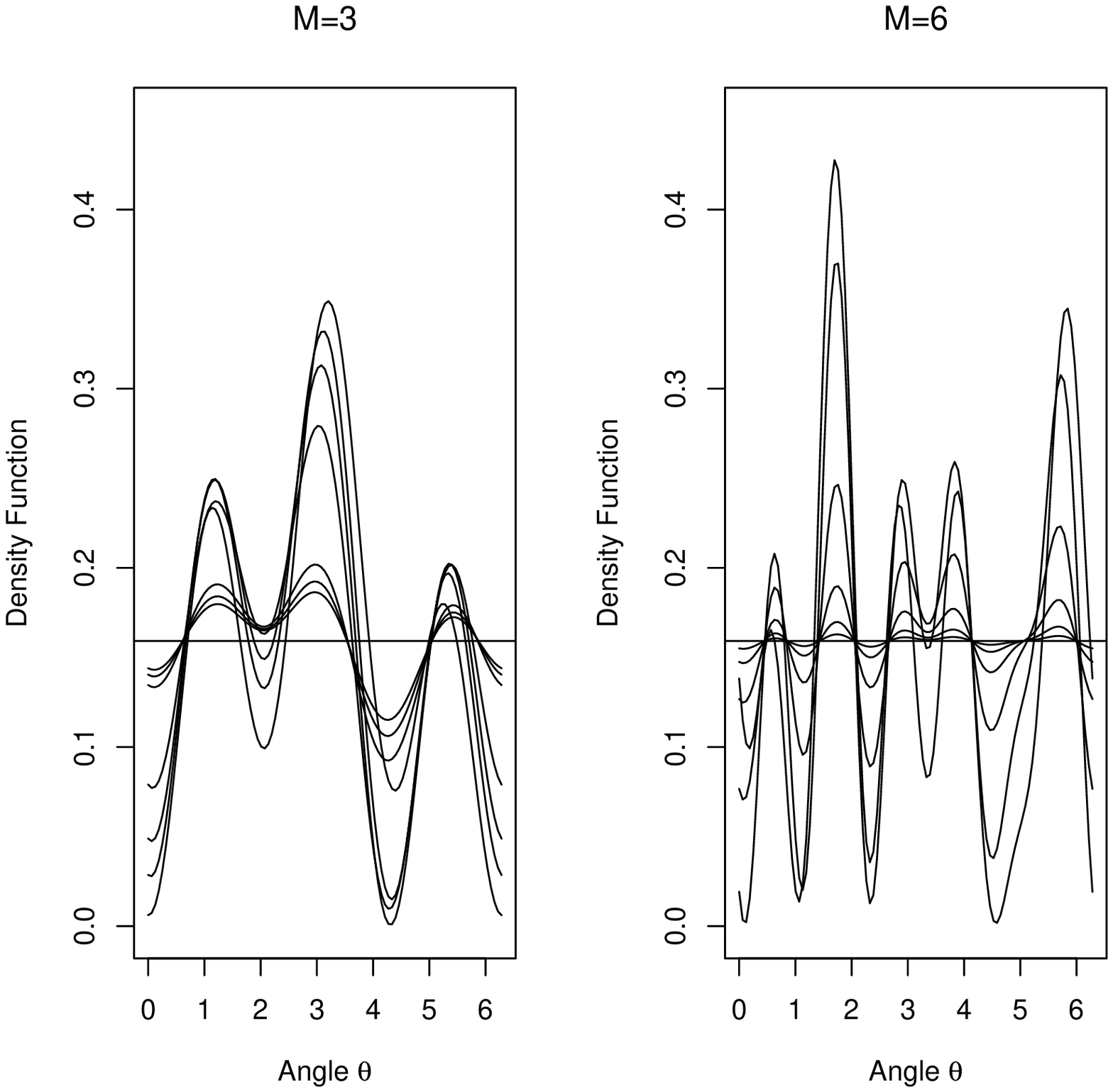}
\end{tabular}
\caption{Cases used for the power and size study for NNTS densities with $M=3$ (left) and $M=6$ (right).
\label{graphpowerexamples}}
\end{table}

\section{Practical Applications}

\subsection{Time of Occurrence of Earthquakes}

In Mexico, three large-intensity earthquakes occurred on September 19, in recent years: in 1985, 2017, and 2022. Moreover, the 2017 and 2022 earthquakes occurred a few minutes after a simulation drill, which is mandatory by law to prepare the general population for this kind of natural phenomenon.
These events have raised concerns
among the public about the fact that a large earthquake occurs randomly with respect to time; thus, it is not possible to predict the specific time at
which an earthquake of high intensity will occur. It is possible to predict the occurrence of replicas of large earthquakes.
We applied the two NNTS ($NNTS1$ and $NNTS2$) and the two Hermans-Rasson and Pycke uniformity tests to test the circular uniformity of the times of occurrence of large
intensity earthquakes since 1900, when more precise instruments to record the time of occurrence of earthquakes became commonly used. The occurrences of
earthquakes with magnitudes greater than 7 (Richter scale) were obtained from the Global Significant Earthquake Database (National Geophysical Data Center/World Data Service (NGDC/WDS): NCEI/WDS Global Significant Earthquake Database). There were a total of 414
earthquakes in the world, and 85 occurred at latitudes from 33.7828 to 8.8243 and longitudes from -118.8281 to -95.2734, which are mainly earthquakes occurring
along the Mexican coast of the Pacific Ocean or in the interior of Mexico. The times of occurrence were transformed into angles by multiplying by $2\pi$ the fraction of
the year in Julian years
at which the earthquake occurred. Figure \ref{graphearthquakesexamples} presents the histograms of the angular values for large
earthquakes occurring worldwide and in
Mexico from 1900 onwards. By applying the $NNTS2$ test with $M=4$, we found that we do not reject the null hypothesis of circular uniformity at a 5\% significance level
with p-values equal to 0.584 for the world earthquakes and 0.635 for the Mexico earthquakes. When using the $NNTS2$ test with $M=3$, the same conclusion was reached
with p-values of 0.366 for the world earthquakes and 0.780 for the Mexico earthquakes. In addition, the modified Hermans-Rasson (p-values of 0.407 and 0.728) and Pycke (p-values
of 0.424 and 0.797) tests did not reject the null hypothesis of circular uniformity. In terms of the analysis in this study, detecting small deviations from uniformity
requires very large sample sizes, and there is no evidence to reject the null hypothesis of circular uniformity with total sample sizes from 1900 onwards.

%earthquakes-2022-09-23_23-59-45_-0500MEXICO8.824285709762464Latitude33.782803497021895.tsv
%maxLongitude=-95.2734375&maxLatitude=33.782803497021895&minLongitude=-118.828125&minLatitude=8.824285709762464

\renewcommand{\tablename}{Figure}
\begin{table}[t]
\centering
\begin{tabular}{c}
%\rotatebox[origin=c]{-90}{
\includegraphics[scale=.75, bb=54 144 558 648]{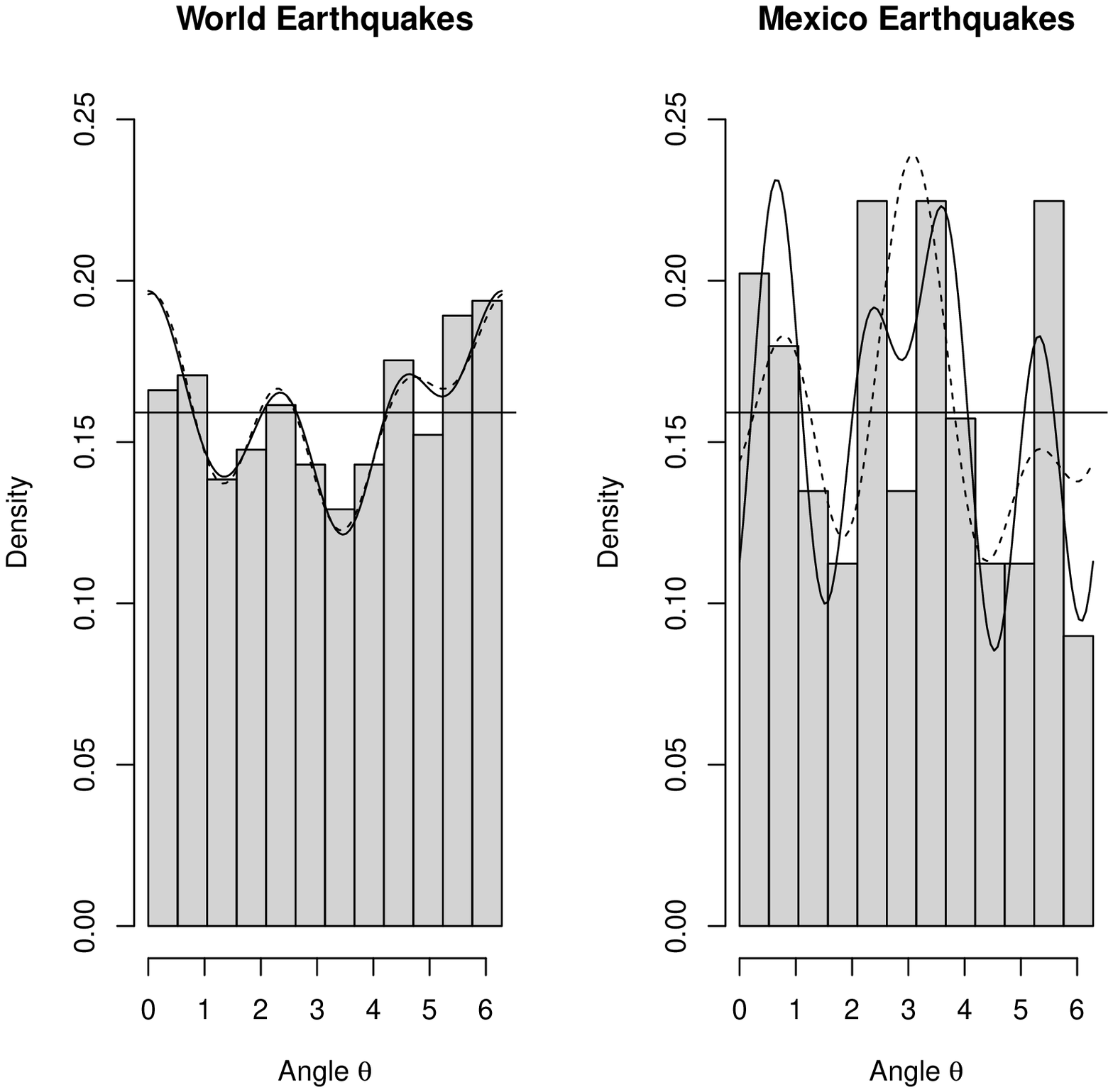}
\end{tabular}
\caption{Earthquakes data: histograms and fitted NNTS densities with $M=3$ and $M=4$ for the angles of occurrence (fraction of the year in Julian years multiplied by $2\pi$) for the world (left) and Mexico (right) earthquakes.
\label{graphearthquakesexamples}}
\end{table}

\subsection{Orientations Taken by Pigeons After Treatment}

Gagliardo et al. (2008) measured the azimuth of vanishing bearings obtained by young homing pigeons randomly assigned to three different groups. The first group (C)
consisted of 41 unmanipulated homing pigeons. The second group (ON) consisted of 27 birds that underwent bilateral olfactory nerve sectioning, and the third
group (V1) included 40 birds that underwent bilateral sectioning of the ophthalmic branch of the trigeminal nerve. The main hypothesis is that, after an intensive training
flight program, pigeons that were deprived of the olfactory nerves (ON) show a circular uniform distribution for their directions of displacement in contrast to the
control (C) and deprived ophthalmic branch of the trigeminal nerve (V1) groups, which show a similar distribution with a preferred direction of displacement. Later, Landler
et al. (2021) considered a subset of the original data of Gagliardo et al. (2008) to test for homogeneity of the circular distributions of the control (25 birds) and
deprived olfactory nerve (25 birds) groups. We applied the Rayleigh, $NNTS2$ with $M=1$ and $M=2$, Hermans-Rasson, and Pycke tests to both datasets, expecting not to reject
the null hypothesis of uniformity for the ON group and to reject the hypothesis of circular uniformity for the C and V1 groups. Table
\ref{pigeonsdata} contains the observed bearings in degrees and the p-values of the different tests applied to each group in both datasets. The Rayleigh test was
implemented because for the C and V1 groups, there appear to be at most two modes (preferred directions). All tests generated p-values in agreement with the
expectation of not rejecting the null hypothesis of circular uniformity for the ON group, in contrast to rejecting it for the C and V1 groups. The only exceptions were
in the reduced dataset of Lander et al. (2021). First, if a researcher considers a significance level equal to 1\%, then only the $NNTS2$ with $M=1$ rejects the null hypothesis of uniformity for the C group, and if a researcher considers a significance level equal to 10\%, then the modified Hermans-Rasson ($HRmT$)
rejects the null hypothesis of uniformity for the ON group.

\section{Conclusions}

Two flexible circular uniformity tests based on maximum likelihood and NNTS multimodal and/or asymmetric distributions were developed as alternative hypotheses, and
their power properties were studied. The null and alternative distributions of the NNTS circular uniformity test statistic are nonstandard asymptotic distributions, and the
common bootstrap procedures are not applicable, given the nonregularity of the maximum likelihood estimator under the null hypothesis of circular uniformity that occurs on
the boundary of the parameter space. Then, the critical
values of the test, or even the p-value, can be obtained by simulation that can be implemented in a reasonable time given the efficient optimisation algorithm developed by
Fern\'{a}ndez-Dur\'{a}n and Gregorio-Dom\'{i}nguez (2010), making the NNTS circular uniformity test suitable for use in practice. The power of the NNTS circular
uniformity test based on the generalised likelihood ratio ($NNTS2$) presents the largest power over the NNTS test based on the standardised maximum likelihood
estimator, $NNTS1$, Pycke test, and modified Hermans-Rasson test in our simulation studies. Then, in circular datasets in which multimodality and/or
asymmetry are present, the NNTS
($NNTS2$) circular uniformity test with an adequate value for parameter $M$ is recommended. In case of doubt regarding the value of the parameter $M$ to use in the NNTS
tests, it is recommended to use the largest value from the set of considered values obtained from theory or from the exploratory inspection of the number of modes in the data. The interpolated critical values for the generalised likelihood ratio NNTS2 test for any sample size were obtained by using regression models that showed an excellent fit.

\section*{Acknowledgements}
The authors wish to thank the Asociaci\'on Mexicana de Cultura, A.C. for its support.

\newpage
\setcounter{table}{0}
\renewcommand{\tablename}{Table}
\renewcommand{\baselinestretch}{1.00}
\begin{table}[t]
\begin{center}
\scalebox{0.85}{
\begin{tabular}{ccccccc}
\hline
    &          & \multicolumn{5}{c}{Sample size} \\
$M$ & $\alpha$ & 25 & 50 & 100 & 200 & 500 \\ 
\hline
1 & 0.10 &  2.7 &  2.5 &  2.4 &  2.3 &  2.3 \\ 
  & 0.05 &  3.7 &  3.3 &  3.1 &  3.1 &  3.1 \\ 
  & 0.01 &  8.0 &  5.3 &  4.8 &  4.8 &  4.7 \\ 
\hline
2 & 0.10 &  8.6 &  4.6 &  4.3 &  4.1 &  3.9 \\ 
  & 0.05 &  9.2 &  5.9 &  5.2 &  5.0 &  4.9 \\ 
  & 0.01 & 10.7 & 10.1 &  7.6 &  7.2 &  6.8 \\ 
\hline
3 & 0.10 &      &  7.9 &  6.1 &  5.6 &  5.4 \\ 
  & 0.05 &      & 13.1 &  7.3 &  6.7 &  6.4 \\ 
  & 0.01 &      & 15.4 & 10.1 &  9.0 &  8.7 \\ 
\hline
4 & 0.10 &      & 12.5 &  7.9 &  7.3 &  6.8 \\ 
  & 0.05 &      & 13.9 &  9.6 &  8.4 &  7.9 \\
  & 0.01 &      & 18.1 & 15.0 & 11.3 & 10.4 \\ 
\hline
5 & 0.10 &      &      & 10.1 &  8.6 &  8.2 \\
  & 0.05 &      &      & 13.0 & 10.0 &  9.4 \\ 
  & 0.01 &      &      & 20.9 & 13.4 & 12.1 \\
\hline
\end{tabular}}
\caption{Critical values for significance levels of 10\%, 5\%, and 1\% and sample sizes of 25, 50, 100, 200, and 500 for the NNTS circular uniformity test based on
the standardised maximum likelihood estimator ($NNTS1$) with test statistic $T_{NNTS1}$. The critical values were obtained from 10000 simulations
from the null circular uniform distribution. For each simulated dataset, the alternative NNTS model is fitted, and the test statistic is calculated.
\label{criticalvalues} }
\end{center}
\end{table}

\newpage
\renewcommand{\tablename}{Table}
\renewcommand{\baselinestretch}{1.00}
\begin{table}[t]
\begin{center}
\scalebox{0.6}{
\begin{tabular}{ccccccccccccccccccccccc}
\hline
    &      & \multicolumn{21}{c}{Sample size}   \\
$M$ & $\alpha$ & 20 & 30 & 40 & 50   & 60   & 70   & 80   & 90   & 100  & 110  & 120  & 130  & 140  & 150  & 200  & 300  & 400  & 500  & 600  & 700  & $\infty$   \\ 
\hline
1 & 0.10 & 5.0  &  4.9 &  4.9 &  4.7 &  4.7 &  4.7 &  4.6 &  4.6 & 4.6  &      &      &      &      &      &      &      &      &      &      &      &  4.6  \\ 
  & 0.05 & 6.6  &  6.3 &  6.3 &  6.1 &  6.1 &  6.1 &  6.1 &  6.1 & 6.1  &      &      &      &      &      &      &      &      &      &      &      &  6.1  \\ 
  & 0.01 & 10.3 &  9.8 &  9.8 &  9.5 &  9.5 &  9.4 &  9.4 &  9.3 & 9.3  &      &      &      &      &      &      &      &      &      &      &      &  9.3  \\ 
\hline
2 & 0.10 &      &  8.5 &  8.3 &  8.1 &  8.1 &  8.0 &  8.0 &  8.0 &  7.9 &  7.9 &      &      &      &      &      &      &      &      &      &      &  7.9  \\ 
  & 0.05 &      & 10.5 & 10.2 &  9.9 &  9.8 &  9.8 &  9.7 &  9.7 &  9.7 &  9.7 &      &      &      &      &      &      &      &      &      &      &  9.7  \\ 
  & 0.01 &      & 14.5 & 14.3 & 14.0 & 13.8 & 13.7 & 13.7 & 13.6 & 13.5 & 13.5 &      &      &      &      &      &      &      &      &      &      & 13.5  \\ 
\hline
3 & 0.10 &      &      & 11.6 & 11.3 & 11.2 & 11.1 & 11.0 & 11.0 & 10.9 & 10.9 & 10.9 & 10.8 & 10.8 & 10.8 & 10.8 & 10.8 &      &      &      &      & 10.8  \\ 
  & 0.05 &      &      & 13.7 & 13.5 & 13.2 & 13.1 & 13.1 & 13.0 & 12.9 & 12.9 & 12.9 & 12.8 & 12.8 & 12.8 & 12.8 & 12.8 &      &      &      &      & 12.8  \\ 
  & 0.01 &      &      & 17.9 & 17.9 & 17.9 & 17.6 & 17.5 & 17.4 & 17.3 & 17.3 & 17.3 & 17.2 & 17.1 & 17.1 & 17.0 & 17.0 &      &      &      &      & 17.0  \\ 
\hline
4 & 0.10 &      &      &      & 14.5 & 14.2 & 14.1 & 13.9 & 13.9 & 13.8 & 13.7 & 13.7 & 13.7 & 13.7 & 13.6 & 13.6 & 13.5 & 13.5 &      &      &      & 13.5  \\ 
  & 0.05 &      &      &      & 16.7 & 16.6 & 16.4 & 16.3 & 16.1 & 16.0 & 15.9 & 15.9 & 15.9 & 15.8 & 15.8 & 15.8 & 15.7 & 15.7 &      &      &      & 15.7  \\ 
  & 0.01 &      &      &      & 21.4 & 21.4 & 21.3 & 20.9 & 20.9 & 20.8 & 20.8 & 20.7 & 20.6 & 20.5 & 20.5 & 20.3 & 20.3 & 20.3 &      &      &      & 20.3  \\ 
\hline
5 & 0.10 &      &      &      &      & 17.3 & 17.1 & 16.9 & 16.8 & 16.7 & 16.6 & 16.5 & 16.5 & 16.4 & 16.3 & 16.3 & 16.2 & 16.1 & 16.1 &      &      & 16.1  \\ 
  & 0.05 &      &      &      &      & 19.7 & 19.6 & 19.5 & 19.2 & 19.1 & 19.0 & 18.9 & 18.9 & 18.8 & 18.7 & 18.6 & 18.6 & 18.5 & 18.5 &      &      & 18.5  \\ 
  & 0.01 &      &      &      &      & 24.6 & 24.6 & 24.5 & 24.4 & 24.3 & 24.2 & 24.0 & 24.0 & 23.8 & 23.6 & 23.6 & 23.5 & 23.4 & 23.4 &      &      & 23.4  \\ 
\hline
6 & 0.10 &      &      &      &      &      & 20.0 & 19.9 & 19.7 & 19.5 & 19.4 & 19.3 & 19.2 & 19.1 & 19.1 & 18.9 & 18.8 & 18.7 & 18.7 & 18.7 &      & 18.7  \\ 
  & 0.05 &      &      &      &      &      & 22.6 & 22.4 & 22.4 & 22.1 & 22.0 & 21.9 & 21.9 & 21.8 & 21.7 & 21.5 & 21.3 & 21.2 & 21.2 & 21.2 &      & 21.2  \\ 
  & 0.01 &      &      &      &      &      & 27.9 & 27.8 & 27.8 & 27.6 & 27.6 & 27.3 & 27.3 & 27.1 & 27.1 & 26.7 & 26.6 & 26.5 & 26.5 & 26.5 &      & 26.5  \\ 
\hline
7 & 0.10 &      &      &      &      &      &      & 22.7 & 22.6 & 22.4 & 22.2 & 22.1 & 22.0 & 21.9 & 21.9 & 21.6 & 21.4 & 21.2 & 21.2 & 21.2 & 21.2 & 21.2  \\ 
  & 0.05 &      &      &      &      &      &      & 25.4 & 25.4 & 25.1 & 25.0 & 24.9 & 24.9 & 24.7 & 24.6 & 24.3 & 24.1 & 24.0 & 24.0 & 23.9 & 23.9 & 23.9  \\ 
  & 0.01 &      &      &      &      &      &      & 31.0 & 31.0 & 30.9 & 30.8 & 30.6 & 30.6 & 30.5 & 30.5 & 29.9 & 29.8 & 29.6 & 29.6 & 29.6 & 29.6 & 29.6  \\ 
\hline
\end{tabular}}
\caption{Critical values for significance levels of 10\%, 5\%, and 1\% and different sample sizes for the NNTS circular uniformity test based on
the generalised likelihood ratio ($NNTS2$) with test statistic $T_{NNTS2}$. The critical values were obtained from 10000 simulations from the null circular uniform
distribution. For each simulated dataset, the alternative NNTS model is fitted, and the test statistic is calculated.
\label{criticalvaluesloglik} }
\end{center}
\end{table}

\newpage

\renewcommand{\tablename}{Table}
\renewcommand{\baselinestretch}{1.00}
\begin{table}[t]
\begin{center}
\scalebox{0.8}{
\begin{tabular}{ccc|c|ccccc|ccc}
\hline
    &           &                  &          & \multicolumn{5}{c|}{Regression Models Estimated Coefficients}                                   &         & abs.  & rel.  \\
$M$ & $SS_{min}$ & $SS_{asymptotic}$ & $\alpha$ & intercept & $M$    & $\frac{1}{SS}$ & $M\left(\frac{1}{SS}\right)$ & $\left(\frac{1}{SS}\right)^2$ & $R^2$   & error & error  \\
\hline
1      &  15    &  85              & 0.10     &   4.5128  &        & 10.8062       &                             &                              & 0.8737  &  0.1  & 2.04\%   \\
       &        &                  & 0.05     &   5.9269  &        & 12.7461       &                             &                              & 0.9229  &  0.1  & 1.59\%   \\
       &        &                  & 0.01     &   9.0630  &        & 24.5377       &                             &                              & 0.9670  &  0.1  & 1.05\%   \\
\hline
2      &  25    &  98              & 0.10     &   7.6807  &        & 24.1698       &                             &                              & 0.9714  &  0.1  & 1.25\%   \\
       &        &                  & 0.05     &   9.3118  &        & 34.1750       &                             &                              & 0.9539  &  0.1  & 1.03\%   \\
       &        &                  & 0.01     &  13.1063  &        & 43.7094       &                             &                              & 0.9791  &  0.1  & 0.70\%   \\
\hline
3 to 7 &        &                  & 0.10     &   3.2703  & 2.5317 & -108.3235     & 32.8331                     & 1618.5535                    & 0.9999  &  0.1  & 0.93\%   \\
       &        &                  & 0.05     &   4.6077  & 2.7291 &  -91.8270     & 31.8820                     & 1368.6187                    & 0.9997  &  0.2  & 1.03\%   \\
       &        &                  & 0.01     &   7.2135  & 3.1555 &   26.9335     & 21.0319                     & -1549.4894                   & 0.9995  &  0.2  & 1.12\%   \\
\hline
3      &  40    &  173             &          &    &    &     &    &      &   &   &   \\
4      &  50    &  203             &          &    &    &     &    &      &   &   &   \\
5      &  60    &  278             &          &    &    &     &    &      &   &   &   \\
6      &  70    &  386             &          &    &    &     &    &      &   &   &   \\
7      &  80    &  562             &          &    &    &     &    &      &   &   &   \\
\hline
\end{tabular}}
\caption{Fitted regression models for the critical values in Table \ref{criticalvaluesloglik} to interpolate critical values of the generalised likelihood ratio test (NNTS2) for any sample size in terms of the reciprocal of the sample size ($SS$), the NNTS parameter $M$ and their interaction and, the reciprocal of the squared sample size. The predictions of the regression models must be rounded to one decimal (precision 0.1). The minimum sample size for which the critical values are valid ($SS_{min}$) and the sample size at which the asymptotic critical values are reached ($SS_{asymptotic}$) are included. Also, the coefficient of determination ($R^2$) of the regression models and the maximum absolute and relative errors of the predicted critical values for the sample sizes in Table \ref{criticalvaluesloglik} are presented.
\label{criticalvaluesregressionmodels} }
\end{center}
\end{table}

\newpage
\renewcommand{\tablename}{Table}
\renewcommand{\baselinestretch}{1.00}
\begin{table}[t]
\begin{center}
\scalebox{0.75}{
\begin{tabular}{cccc|ccccc|ccccc|ccc}
\hline
     &    &       &  & \multicolumn{5}{|c|}{$NNTS1$  std. max. lik. est.} & \multicolumn{5}{|c|}{$NNTS2$ Likelihood Ratio} & & & \\
\hline
case & SS & $c_0$ & $1-c_0^2$  & 1 & 2 & 3 & 4 & 5 & 1 & 2 & 3 & 4 & 5 & RT & HRmT  & PT  \\
\hline
1 & 25   & 0.59 & 0.6519 & 46 & 33 &    &    &    & 46 & 35 &  &  &  & 46 & 41 & 56 \\
2 & 25   & 0.67 & 0.5511 & 37 & 30 &    &    &    & 39 & 32 &  &  &  & 41 & 37 & 55 \\
3 & 25   & 0.75 & 0.4375 & 30 & 35 &    &    &    & 33 & 33 &  &  &  & 36 & 37 & 55 \\
4 & 25   & 0.83 & 0.3111 & 29 & 38 &    &    &    & 33 & 37 &  &  &  & 36 & 37 & 53 \\
5 & 25   & 0.91 & 0.1719 & 28 & 34 &    &    &    & 30 & 32 &  &  &  & 30 & 30 & 40 \\
6 & 25 & 0.9899 & 0.0201 & 14 & 13 &    &    &    & 13 & 15 &  &  &  & 15 & 15 & 16 \\
7 & 25 & 0.9939 & 0.0122 & 12 & 13 &    &    &    & 12 & 13 &  &  &  & 12 & 11 & 12 \\
8 & 25 & 0.9959 & 0.0082 & 10 & 11 &    &    &    & 10 & 11 &  &  &  & 11 & 11 & 11 \\
\hline
1 & 50   & 0.59 & 0.6519 & 71 & 63 & \underline{92} & 85              &  & 72 & 61 & \underline{95} & \underline{92} &  & 71 & 67 & 89 \\
2 & 50   & 0.67 & 0.5511 & 64 & 57 & \underline{92} & 87              &  & 65 & 58 & \underline{95} & \underline{93} &  & 67 & 67 & 89 \\
3 & 50   & 0.75 & 0.4375 & 55 & 62 & \underline{91} & 85              &  & 59 & 56 & \underline{96} & \underline{93} &  & 63 & 63 & 88 \\
4 & 50   & 0.83 & 0.3111 & 50 & 72 & \underline{95} & \underline{90}  &  & 53 & 63 & \underline{96} & \underline{93} &  & 57 & 64 & 87 \\
5 & 50   & 0.91 & 0.1719 & 43 & 67 & \underline{83} & \underline{75}  &  & 46 & 59 & \underline{86} & \underline{80} &  & 47 & 55 & 70 \\
6 & 50 & 0.9899 & 0.0201 & 17 & 19 &            17  & 17              &  & 18 & 19 &            20  &            18  &  & 18 & 18 & 19 \\
7 & 50 & 0.9939 & 0.0122 & 12 & 14 &            15  & 16              &  & 12 & 14 &            17  &            17  &  & 12 & 14 & 16 \\
8 & 50 & 0.9959 & 0.0082 & 12 & 13 &            13  & 13              &  & 12 & 12 &            14  &            13  &  & 12 & 11 & 12 \\
\hline
1 & 100   & 0.59 & 0.6519 & 96 & 90 & 100 & 100 & 100 & 96 & 90 & 100 & 100 & 100 & 95 & 95 & 100 \\
2 & 100   & 0.67 & 0.5511 & 90 & 84 & 100 & 100 & 100 & 91 & 84 & 100 & 100 & 100 & 91 & 93 & 100 \\
3 & 100   & 0.75 & 0.4375 & 87 & 88 & 100 & 100 & 100 & 89 & 86 & 100 & 100 & 100 & 90 & 93 & 100 \\
4 & 100   & 0.83 & 0.3111 & 83 & 95 & 100 & 100 & 100 & 86 & 92 & 100 & 100 & 100 & 87 & 94 & 100 \\
5 & 100   & 0.91 & 0.1719 & 74 & 93 & 100 &  99 &  99 & 76 & 89 & 100 &  99 &  99 & 78 & 90 &  98 \\
6 & 100 & 0.9899 & 0.0201 & 25 & 29 &  \underline{35} &  32 &  28 & 25 & 28 &  \underline{35} &  33 &  31 & 24 & 27 &  31 \\
7 & 100 & 0.9939 & 0.0122 & 20 & 21 &  24 &  21 &  21 & 20 & 20 &  24 &  21 &  22 & 19 & 21 &  23 \\
8 & 100 & 0.9959 & 0.0082 & 17 & 19 &  20 &  19 &  18 & 17 & 19 &  21 &  21 &  18 & 17 & 19 &  20 \\
\hline
1 & 200   & 0.59 & 0.6519 & 100 & 100 & 100 & 100 & 100 & 100 & 100 & 100 & 100 & 100 & 100 & 100 & 100 \\
2 & 200   & 0.67 & 0.5511 &  99 &  99 & 100 & 100 & 100 & 100 &  99 & 100 & 100 & 100 & 100 & 100 & 100 \\
3 & 200   & 0.75 & 0.4375 & 100 &  99 & 100 & 100 & 100 & 100 &  99 & 100 & 100 & 100 & 100 & 100 & 100 \\
4 & 200   & 0.83 & 0.3111 &  98 & 100 & 100 & 100 & 100 &  99 & 100 & 100 & 100 & 100 &  99 & 100 & 100 \\
5 & 200   & 0.91 & 0.1719 &  95 & 100 & 100 & 100 & 100 &  95 &  99 & 100 & 100 & 100 &  95 & 100 & 100 \\
6 & 200 & 0.9899 & 0.0201 &  40 &  50 &  \underline{63} &  57 &  55 &  41 &  47 &  \underline{62} &  57 &  53 &  40 &  48 &  57 \\
7 & 200 & 0.9939 & 0.0122 &  31 &  34 &  \underline{43} &  37 &  37 &  31 &  33 &  \underline{43} &  37 &  36 &  30 &  33 &  39 \\
8 & 200 & 0.9959 & 0.0082 &  24 &  28 &  \underline{32} &  28 &  27 &  23 &  27 &  31 &  29 &  27 &  23 &  26 &  29 \\
\hline
1 & 500 & 0.59   & 0.6519 & 100 & 100 & 100 & 100 & 100 & 100 & 100 & 100 & 100 & 100 & 100 & 100 & 100 \\
2 & 500 & 0.67   & 0.5511 & 100 & 100 & 100 & 100 & 100 & 100 & 100 & 100 & 100 & 100 & 100 & 100 & 100 \\
3 & 500 & 0.75   & 0.4375 & 100 & 100 & 100 & 100 & 100 & 100 & 100 & 100 & 100 & 100 & 100 & 100 & 100 \\
4 & 500 & 0.83   & 0.3111 & 100 & 100 & 100 & 100 & 100 & 100 & 100 & 100 & 100 & 100 & 100 & 100 & 100 \\
5 & 500 & 0.91   & 0.1719 & 100 & 100 & 100 & 100 & 100 & 100 & 100 & 100 & 100 & 100 & 100 & 100 & 100 \\
6 & 500 & 0.9899 & 0.0201 &  76 &  87 &  97 &  96 &  94 &  75 &  84 &  97 &  96 &  94 &  74 &  87 &  95 \\
7 & 500 & 0.9939 & 0.0122 &  54 &  66 &  \underline{82} &  \underline{79} &  75 &  53 &  65 &  \underline{82} &  78 &  74 &  53 &  65 &  76 \\
8 & 500 & 0.9959 & 0.0082 &  45 &  54 &  \underline{68} &             63  &  58 &  46 &  52 &  \underline{67} &  62 &  57 &  46 &  52 &  61 \\
\hline
\end{tabular}}
\caption{Power comparison of the $NNTS1$ with $M=1, 2, 3, 4, \mbox{ and } 5$, $NNTS2$ with $M=1, 2, 3, 4, \mbox{ and } 5$, Rayleigh test ($RT$), modified Hermans-Rasson
($HRmT$), and Pycke ($PT$) tests considering a significance level
$\alpha=10\%$. The power of the tests is obtained from the simulation of 1000 datasets from an NNTS density with $M=3$ (see left plot in Figure \ref{graphpowerexamples}),
applying the various tests to each of the datasets and, calculating the frequency at which the null hypothesis of circular uniformity is rejected. Underlined numbers are
examples for which the $NNTS1$ or $NNTS2$ power is greater than the Pycke power by at least 3 (3\%).
\label{powervalues10} }
\end{center}
\end{table}

\newpage
\renewcommand{\tablename}{Table}
\renewcommand{\baselinestretch}{1.00}
\begin{table}[t]
\begin{center}
\scalebox{0.75}{
\begin{tabular}{cccc|ccccc|ccccc|ccc}
\hline
     &    &       &  & \multicolumn{5}{|c|}{$NNTS1$  std. max. lik. est.} & \multicolumn{5}{|c|}{$NNTS2$ Likelihood Ratio} & & & \\
\hline
case & SS & $c_0$ & $1-c_0^2$  & 1 & 2 & 3 & 4 & 5 & 1 & 2 & 3 & 4 & 5 & RT & HRmT  & PT  \\
\hline
1 & 25 & 0.59   & 0.6519 & 32 & 24 &  &  &  & 33 & 21 &  &  &  & 32 & 28 & 43 \\
2 & 25 & 0.67   & 0.5511 & 25 & 24 &  &  &  & 27 & 21 &  &  &  & 30 & 25 & 41 \\
3 & 25 & 0.75   & 0.4375 & 20 & 26 &  &  &  & 22 & 23 &  &  &  & 24 & 26 & 40 \\
4 & 25 & 0.83   & 0.3111 & 18 & 27 &  &  &  & 22 & 24 &  &  &  & 25 & 26 & 37 \\
5 & 25 & 0.91   & 0.1719 & 17 & 23 &  &  &  & 18 & 19 &  &  &  & 19 & 19 & 27 \\
6 & 25 & 0.9899 & 0.0201 &  8 &  8 &  &  &  &  8 &  9 &  &  &  &  8 &  9 &  9 \\
7 & 25 & 0.9939 & 0.0122 &  6 &  7 &  &  &  &  6 &  6 &  &  &  &  6 &  5 &  6 \\
8 & 25 & 0.9959 & 0.0082 &  5 &  5 &  &  &  &  5 &  5 &  &  &  &  5 &  5 &  6 \\
\hline
1 & 50 & 0.59    & 0.6519 & 58 & 50 & 80              & 76              &     & 59 & 49 & \underline{89}  & \underline{85}  &     & 58 & 56 & 80 \\
2 & 50 & 0.67    & 0.5511 & 50 & 46 & 76              & 78              &     & 54 & 45 & \underline{92}  & \underline{88}  &     & 55 & 53 & 80 \\
3 & 50 & 0.75    & 0.4375 & 40 & 48 & 76              & 77              &     & 46 & 44 & \underline{91}  & \underline{86}  &     & 49 & 50 & 79 \\
4 & 50 & 0.83    & 0.3111 & 35 & 60 & \underline{84}  & 80              &     & 40 & 50 & \underline{91}  & \underline{87}  &     & 43 & 50 & 76 \\
5 & 50 & 0.91    & 0.1719 & 30 & 56 & \underline{69}  & \underline{61}  &     & 31 & 47 & \underline{75}  & \underline{68}  &     & 34 & 40 & 57 \\
6 & 50 & 0.9899  & 0.0201 & 10 & 11 & 10              & 10              &     &  9 & 11 & 11              & 10              &     &  9 & 10 & 11 \\
7 & 50 & 0.9939  & 0.0122 &  7 &  8 &  9              &  9              &     &  6 &  6 &  9              &  9              &     &  6 &  6 &  8 \\
8 & 50 & 0.9959  & 0.0082 &  6 &  8 &  7              &  6              &     &  6 &  7 &  7              &  7              &     &  6 &  6 &  7 \\
\hline
1 & 100 & 0.59   & 0.6519 & 92 & 83 & 100             & 100             &  99 & 92 & 84 & 100             & 100             & 100 & 91 & 92 & 100 \\
2 & 100 & 0.67   & 0.5511 & 83 & 75 & 100             & 100             &  98 & 85 & 76 & 100             & 100             & 100 & 84 & 86 & 100 \\
3 & 100 & 0.75   & 0.4375 & 76 & 82 & 100             & 100             &  99 & 81 & 79 & 100             & 100             & 100 & 83 & 87 &  99 \\
4 & 100 & 0.83   & 0.3111 & 71 & 91 & 100             & 100             & 100 & 76 & 85 & 100             & 100             & 100 & 79 & 88 &  99 \\
5 & 100 & 0.91   & 0.1719 & 63 & 89 &  \underline{99} &  98             &  95 & 66 & 83 &  \underline{99} &  \underline{99} &  97 & 67 & 82 &  96 \\
6 & 100 & 0.9899 & 0.0201 & 15 & 18 &  \underline{23} &  20             &  13 & 16 & 17 &  \underline{24} &  20             &  20 & 16 & 16 &  19 \\
7 & 100 & 0.9939 & 0.0122 & 12 & 12 &  14             &  12             &  11 & 12 & 11 &  15             &  12             &  14 & 12 & 12 &  13 \\
8 & 100 & 0.9959 & 0.0082 & 11 & 13 &  13             &  11             &   7 & 10 & 12 &  14             &  11             &  11 & 11 & 12 &  12 \\
\hline
1 & 200 & 0.59   & 0.6519 & 100 &  99 & 100 & 100 & 100 & 100 & 99 & 100 & 100 & 100 & 100 & 100 & 100 \\
2 & 200 & 0.67   & 0.5511 &  98 &  98 & 100 & 100 & 100 &  99 & 98 & 100 & 100 & 100 &  99 & 100 & 100 \\
3 & 200 & 0.75   & 0.4375 &  98 &  99 & 100 & 100 & 100 &  98 & 99 & 100 & 100 & 100 &  99 & 100 & 100 \\
4 & 200 & 0.83   & 0.3111 &  96 & 100 & 100 & 100 & 100 &  97 & 99 & 100 & 100 & 100 &  97 & 100 & 100 \\
5 & 200 & 0.91   & 0.1719 &  90 & 100 & 100 & 100 & 100 &  91 & 98 & 100 & 100 & 100 &  91 &  99 & 100 \\
6 & 200 & 0.9899 & 0.0201 &  27 &  37 &  \underline{50} &  44 &  41 &  27 & 35 &  \underline{50} &  44 &  40 &  27 &  34 &  43 \\
7 & 200 & 0.9939 & 0.0122 &  19 &  24 &  \underline{30} &  26 &  25 &  19 & 23 &  \underline{29} &  26 &  24 &  19 &  22 &  25 \\
8 & 200 & 0.9959 & 0.0082 &  15 &  19 &  21 &  18 &  17 &  15 & 18 &  20 &  18 &  17 &  15 &  16 &  19 \\
\hline
1 & 500 & 0.59   & 0.6519 & 100 & 100 & 100 & 100 & 100 & 100 & 100 & 100 & 100 & 100 & 100 & 100 & 100 \\
2 & 500 & 0.67   & 0.5511 & 100 & 100 & 100 & 100 & 100 & 100 & 100 & 100 & 100 & 100 & 100 & 100 & 100 \\
3 & 500 & 0.75   & 0.4375 & 100 & 100 & 100 & 100 & 100 & 100 & 100 & 100 & 100 & 100 & 100 & 100 & 100 \\
4 & 500 & 0.83   & 0.3111 & 100 & 100 & 100 & 100 & 100 & 100 & 100 & 100 & 100 & 100 & 100 & 100 & 100 \\
5 & 500 & 0.91   & 0.1719 & 100 & 100 & 100 & 100 & 100 & 100 & 100 & 100 & 100 & 100 & 100 & 100 & 100 \\
6 & 500 & 0.9899 & 0.0201 &  64 &  79 &  \underline{95} &  92 &  88 &  63 &  76 &  \underline{95} &  92 &  87 &  63 &  77 &  90 \\
7 & 500 & 0.9939 & 0.0122 &  41 &  56 &  \underline{73} &  \underline{70} &  64 &  41 &  53 &  \underline{72} &  \underline{69} &  64 &  41 &  54 &  66 \\
8 & 500 & 0.9959 & 0.0082 &  32 &  39 &  \underline{55} &  \underline{50} &  45 &  32 &  37 &  \underline{55} &  \underline{50} &  44 &  32 &  38 &  47 \\
\hline
\end{tabular}}
\caption{Power comparison of the $NNTS1$ with $M=1, 2, 3, 4, \mbox{ and } 5$, $NNTS2$ with $M=1, 2, 3, 4, \mbox{ and } 5$, Rayleigh test ($RT$), modified Hermans-Rasson
($HRmT$), and Pycke ($PT$) tests considering a significance level
$\alpha=5\%$. The power of the tests is obtained from the simulation of 1000 datasets from an NNTS density with $M=3$ (see left plot in Figure \ref{graphpowerexamples}),
applying the various tests to each of the datasets and, calculating the frequency at which the null hypothesis of circular uniformity is rejected. Underlined numbers are
examples for which the $NNTS1$ or $NNTS2$ power is greater than the Pycke power by at least 3 (3\%).
\label{powervalues05} }
\end{center}
\end{table}

\newpage
\renewcommand{\tablename}{Table}
\renewcommand{\baselinestretch}{1.00}
\begin{table}[t]
\begin{center}
\scalebox{0.75}{
\begin{tabular}{cccc|ccccc|ccccc|ccc}
\hline
     &    &       &  & \multicolumn{5}{|c|}{$NNTS1$ std. max. lik. est.} & \multicolumn{5}{|c|}{$NNTS2$ Likelihood Ratio} & & & \\
\hline
case & SS & $c_0$ & $1-c_0^2$  & 1 & 2 & 3 & 4 & 5 & 1 & 2 & 3 & 4 & 5 & RT & HRmT  & PT  \\
\hline
1 & 25 & 0.59   & 0.6519 & 9 & 10 &  &  &  & 12 & 9 &  &  &  & 14 & 12 & 20 \\
2 & 25 & 0.67   & 0.5511 & 7 &  9 &  &  &  & 10 & 7 &  &  &  & 12 & 10 & 17 \\
3 & 25 & 0.75   & 0.4375 & 6 & 11 &  &  &  &  9 & 9 &  &  &  & 11 & 11 & 17 \\
4 & 25 & 0.83   & 0.3111 & 4 & 11 &  &  &  &  8 & 8 &  &  &  & 10 &  9 & 15 \\
5 & 25 & 0.91   & 0.1719 & 5 &  7 &  &  &  &  6 & 6 &  &  &  &  6 &  5 &  8 \\
6 & 25 & 0.9899 & 0.0201 & 2 &  2 &  &  &  &  2 & 2 &  &  &  &  2 &  1 &  1 \\
7 & 25 & 0.9939 & 0.0122 & 2 &  2 &  &  &  &  1 & 1 &  &  &  &  1 &  1 &  1 \\
8 & 25 & 0.9959 & 0.0082 & 1 &  1 &  &  &  &  1 & 1 &  &  &  &  1 &  1 &  1 \\
\hline
1 & 50 & 0.59   & 0.6519 & 35 & 23 & 53             & 41 &  & 36 & 26 & \underline{74} & \underline{65} &  & 35 & 31 & 54 \\
2 & 50 & 0.67   & 0.5511 & 25 & 17 & 56             & 45 &  & 29 & 24 & \underline{74} & \underline{66} &  & 31 & 29 & 55 \\
3 & 50 & 0.75   & 0.4375 & 18 & 20 & \underline{56} & 48 &  & 23 & 25 & \underline{71} & \underline{63} &  & 26 & 28 & 50 \\
4 & 50 & 0.83   & 0.3111 & 14 & 31 & \underline{49} & 44 &  & 20 & 28 & \underline{73} & \underline{64} &  & 22 & 22 & 46 \\
5 & 50 & 0.91   & 0.1719 & 13 & 28 & 31             & 25 &  & 15 & 24 & \underline{49} & \underline{40} &  & 16 & 17 & 29 \\
6 & 50 & 0.9899 & 0.0201 &  2 &  3 &  3             &  2 &  &  2 &  4 &  3             &  3             &  &  2 &  3 &  2 \\
7 & 50 & 0.9939 & 0.0122 &  2 &  2 &  2             &  2 &  &  2 &  2 &  3             &  3             &  &  2 &  2 &  1 \\
8 & 50 & 0.9959 & 0.0082 &  2 &  2 &  2             &  2 &  &  1 &  2 &  2             &  1             &  &  2 &  1 &  1 \\
\hline
1 & 100 & 0.59   & 0.6519 & 79 & 66 &  \underline{99} & 96 & 89 & 77 & 64 &  \underline{99} & \underline{99} & 98 & 75 & 74 & 96 \\
2 & 100 & 0.67   & 0.5511 & 62 & 52 &  \underline{99} & 95 & 84 & 65 & 53 &  \underline{99} & \underline{99} & 97 & 66 & 68 & 94 \\
3 & 100 & 0.75   & 0.4375 & 51 & 62 &  \underline{99} & 96 & 87 & 58 & 60 & \underline{100} & \underline{99} & \underline{98} & 62 & 67 & 95 \\
4 & 100 & 0.83   & 0.3111 & 46 & 77 & \underline{100} & \underline{98} & 94 & 51 & 67 & \underline{100} & \underline{99} & \underline{99} & 56 & 68 & 95 \\
5 & 100 & 0.91   & 0.1719 & 39 & 72 &  \underline{96} & \underline{87} & 79 & 39 &  6 &  \underline{96} & \underline{94} & \underline{90} & 39 & 54 & 82 \\
6 & 100 & 0.9899 & 0.0201 &  5 &  7 &   \underline{9} &  5 &  5 &  5 &  5 &   \underline{9} &  8 &  5 &  5 &  5 &  6 \\
7 & 100 & 0.9939 & 0.0122 &  3 &  4 &   4 &  3 &  3 &  3 &  4 &   5 &  3 &  4 &  3 &  3 &  4 \\
8 & 100 & 0.9959 & 0.0082 &  3 &  4 &   4 &  2 &  2 &  4 &  4 &   4 &  4 &  2 &  3 &  4 &  3 \\
\hline
1 & 200 & 0.59   & 0.6519 & 98 & 97 & 100 & 100 & 100 & 99 & 98 & 100 & 100 & 100 & 98 & 99 & 100 \\
2 & 200 & 0.67   & 0.5511 & 95 & 92 & 100 & 100 & 100 & 96 & 93 & 100 & 100 & 100 & 96 & 98 & 100 \\
3 & 200 & 0.75   & 0.4375 & 91 & 96 & 100 & 100 & 100 & 93 & 95 & 100 & 100 & 100 & 94 & 99 & 100 \\
4 & 200 & 0.83   & 0.3111 & 86 & 99 & 100 & 100 & 100 & 89 & 97 & 100 & 100 & 100 & 91 & 97 & 100 \\
5 & 200 & 0.91   & 0.1719 & 74 & 97 & 100 & 100 & 100 & 77 & 94 & 100 & 100 & 100 & 78 & 94 & 100 \\
6 & 200 & 0.9899 & 0.0201 & 12 & 16 &  \underline{28} &  21 &  17 & 11 & 15 &  \underline{28} &  \underline{23} &  19 & 12 & 14 &  19 \\
7 & 200 & 0.9939 & 0.0122 &  7 &  9 &  \underline{13} &  10 &   8 &  7 &  8 &  \underline{14} &  \underline{12} &   9 &  7 &  7 &   8 \\
8 & 200 & 0.9959 & 0.0082 &  5 &  5 &   \underline{8} &   5 &   4 &  5 &  4 &   \underline{8} &   6 &   5 &  5 &  4 &   5 \\
\hline
1 & 500 & 0.59   & 0.6519 & 100 & 100 & 100 & 100 & 100 & 100 & 100 & 100 & 100 & 100 & 100 & 100 & 100 \\
2 & 500 & 0.67   & 0.5511 & 100 & 100 & 100 & 100 & 100 & 100 & 100 & 100 & 100 & 100 & 100 & 100 & 100 \\
3 & 500 & 0.75   & 0.4375 & 100 & 100 & 100 & 100 & 100 & 100 & 100 & 100 & 100 & 100 & 100 & 100 & 100 \\
4 & 500 & 0.83   & 0.3111 & 100 & 100 & 100 & 100 & 100 & 100 & 100 & 100 & 100 & 100 & 100 & 100 & 100 \\
5 & 500 & 0.91   & 0.1719 & 100 & 100 & 100 & 100 & 100 & 100 & 100 & 100 & 100 & 100 & 100 & 100 & 100 \\
6 & 500 & 0.9899 & 0.0201 &  38 &  60 &  \underline{82} &  \underline{76} &  \underline{72} &  37 &  55 &  \underline{80} &  \underline{74} &  71 &  38 &  55 &  69 \\
7 & 500 & 0.9939 & 0.0122 &  21 &  34 &  \underline{49} &  \underline{42} &  37 &  21 &  31 &  \underline{48} &  \underline{41} &  38 &  22 &  30 &  38 \\
8 & 500 & 0.9959 & 0.0082 &  13 &  21 &  \underline{31} &  25 &  22 &  13 &  18 &  \underline{30} &  25 &  22 &  13 &  18 &  23 \\
\hline
\end{tabular}}
\caption{Power comparison of the $NNTS1$ with $M=1, 2, 3, 4, \mbox{ and } 5$, $NNTS2$ with $M=1, 2, 3, 4, \mbox{ and } 5$, Rayleigh test ($RT$), modified Hermans-Rasson
($HRmT$), and Pycke ($PT$) tests considering a significance level
$\alpha=1\%$. The power of the tests are obtained from the simulation of 1000 datasets from an NNTS density with $M=3$ (see left plot in Figure \ref{graphpowerexamples}),
applying the various tests to each of the datasets and, calculating the frequency at which the null hypothesis of circular uniformity is rejected. Underlined numbers
are examples for which the $NNTS1$ or $NNTS2$ power is greater than the Pycke power by at least 3 (3\%).
\label{powervalues01} }
\end{center}
\end{table}

\newpage
\renewcommand{\tablename}{Table}
\renewcommand{\baselinestretch}{1.00}
\begin{table}[t]
\begin{center}
\scalebox{0.75}{
\begin{tabular}{c|c|ccc|ccc|ccc}
\hline
case  &       & \multicolumn{3}{|c|}{$\alpha=0.10$} & \multicolumn{3}{|c|}{$\alpha=0.05$} & \multicolumn{3}{|c}{$\alpha=0.01$} \\
$c_0$ &  Test & 100 & 200 & 500 & 100 & 200 & 500 & 100 & 200 & 500 \\
\hline
1         & $M=6$ & 100 & 100 & 100 & 100 & 100 & 100 &  \underline{100} & 100 & 100 \\
0.5072892 & $M=7$ & 100 & 100 & 100 & 100 & 100 & 100 &  \underline{100} & 100 & 100 \\
          & Pycke & 100 & 100 & 100 &  99 & 100 & 100 &   96             & 100 & 100 \\
\hline
2         & $M=6$  & \underline{100} & 100 & 100 & \underline{100} & 100 & 100 &  \underline{99} & 100 & 100 \\
0.8594613 & $M=7$  & \underline{100} & 100 & 100 & \underline{100} & 100 & 100 &  \underline{98} & 100 & 100 \\
          & Pycke  & 97              & 100 & 100 &  92             & 100 & 100 &    69           & 100 & 100 \\
\hline
3         & $M=6$ & \underline{52} & \underline{85} & 100 & \underline{41} & \underline{74} & \underline{100} &  \underline{16} & \underline{50} & \underline{99} \\
0.9789961 & $M=7$ & \underline{50} & \underline{82} & 100 & \underline{37} & \underline{71} & \underline{100} &  \underline{14} & \underline{46} & \underline{98} \\
          & Pycke & 34             & 58             &  99 & 22             & 43             &  95             &  6              & 20             & 78             \\
\hline
4         & $M=6$  & 14 &\underline{21} & \underline{39}  & 8 & 10 & \underline{26} &  3 & 3 & \underline{12} \\
0.9973522 & $M=7$  & 15 & 17            & \underline{37}  & 8 & 10 & \underline{26} &  2 & 3 & \underline{10} \\
          & Pycke  & 14 & 15            & 25              & 8 &  8 & 16             &  2 & 1 &  4             \\
\hline
5         & $M=6$  & 11 & 10 & 13  & 6 & 5 & 6 &  1 & 1 & 2 \\
0.9996743 & $M=7$  & 11 & 10 & 14  & 5 & 5 & 7 &  1 & 1 & 2 \\
          & Pycke  & 11 & 10 & 12  & 5 & 5 & 6 &  1 & 1 & 2 \\
\hline
6         & $M=6$  & 11 & 10 & 11  & 6 & 5 & 6 &  1 & 1 & 2 \\
0.9999601 & $M=7$  & 10 & 10 & 11  & 6 & 5 & 6 &  1 & 1 & 1 \\
          & Pycke  & 11 &  9 & 11  & 5 & 5 & 5 &  1 & 1 & 1 \\
\hline
\end{tabular}}
\caption{Power comparison of the likelihood ratio $NNTS2$ with $M=6$, likelihood ratio $NNTS2$ with $M=7$, and Pycke tests considering significance levels $\alpha$ = 10\%,
5\%, and 1\%. The power of the tests are obtained from the simulation of 1000 datasets from an NNTS density with $M=6$ (see right plot in Figure
\ref{graphpowerexamples}), applying the various tests to each of the datasets and, calculating the frequency at which the null hypothesis of circular uniformity is
rejected. Underlined numbers are examples for which the $NNTS1$ or $NNTS2$ power is greater than the Pycke test power by at least 3 (3\%).
\label{powervaluesM6} }
\end{center}
\end{table}

\newpage
\renewcommand{\tablename}{Table}
\renewcommand{\baselinestretch}{1.00}
\begin{table}[t]
\begin{center}
\scalebox{0.75}{
\begin{tabular}{c|c|ccccc}
\hline
\multicolumn{2}{c|}{Reduced Dataset in Landler et al. (2021)} & \multicolumn{5}{|c}{Test p-value} \\
\hline
Group & Observed bearings (azimuth) in degrees & $RT$ & $HRmT$ & $PT$ & $M=1$  & $M=2$  \\
\hline
%5, 20, 45, 50, 145, 170, 205, 210, 210, 210, 215, 230 ,230, 240, 240, 270, 270, 300, 310, 310, 310, 320, 330, 340, 350
C      & 5, 20, 45, 50, 145, 170, 205, 210, 210, 210, 215, 230 ,230,  & 0.017  & 0.032  & 0.031  & 0.006  & 0.022   \\
(25)   & 240, 240, 270, 270, 300, 310, 310, 310, 320, 330, 340, 350   &        &        &        & 11.26  & 12.53   \\
\hline
%20, 40, 45, 50, 60, 60, 60, 70, 80, 90, 90, 90, 110, 130, 140, 170, 210, 210, 215, 230, 270, 270, 295, 320, 325
ON     & 20, 40, 45, 50, 60, 60, 60, 70, 80, 90, 90, 90, 110, 130,  & 0.222  & 0.078  & 0.125  & 0.321  & 0.175   \\
(25)   & 140, 170, 210, 210, 215, 230, 270, 270, 295, 320, 325      &        &        &        & 2.42   & 6.96    \\
\hline
\multicolumn{2}{c}{Complete Dataset in Gagliardo et al. (2008)} & \multicolumn{5}{|c}{Test p-value} \\
\hline
Group & Observed bearings (azimuth) in degrees & $RT$ & $HRmT$ & $PT$ & $M=1$  & $M=2$  \\
\hline
%1, 2, 3, 8, 10, 10, 10, 12, 14, 18, 18, 19, 42, 46, 46,
%48, 52, 54, 58, 86, 92, 108, 131, 274, 306, 310, 320, 324, 327, 328,
%333, 334, 334, 336, 342, 346, 350, 350, 352, 354, 358
C      & 1, 2, 3, 8, 10, 10, 10, 12, 14, 18, 18, 19, 42, 46, 46, 48, & 0.000  & 0.000  & 0.000  & 0.000  & 0.000   \\
(41)   & 52, 54, 58, 86, 92, 108, 131, 274, 306, 310, 320, 324, 327, &        &        &        & 43.10  & 53.75   \\
       & 328, 333, 334, 334, 336, 342, 346, 350, 350, 352, 354, 358  &        &        &        &        &         \\
\hline
%4, 11, 38, 47, 52, 79, 106, 106, 120, 126, 138, 142, 146, 154, 158,
%182, 194, 252, 268, 292, 292, 298, 308, 323, 324, 338, 344
ON     & 4, 11, 38, 47, 52, 79, 106, 106, 120, 126, 138, 142, 146, 154,   & 0.796  & 0.275  & 0.598  & 0.725  & 0.170   \\
(27)   & 158, 182, 194, 252, 268, 292, 292, 298, 308, 323, 324, 338, 344  &        &        &        & 0.69   & 7.08    \\
\hline
%3, 4, 4, 4, 6, 6, 8, 16, 17, 21, 22, 24, 24, 40, 44,
%46, 70, 80, 81, 84, 88, 102, 124, 267, 294, 304, 322, 334, 336, 338,
%339, 342, 344, 349, 353, 354, 354, 356, 358, 358
V1     & 3, 4, 4, 4, 6, 6, 8, 16, 17, 21, 22, 24, 24, 40, 44, 46, 70,  & 0.000  & 0.000  & 0.000  & 0.000  & 0.000   \\
(40)   & 80, 81, 84, 88, 102, 124, 267, 294, 304, 322, 334, 336,       &        &        &        & 41.80  & 51.82   \\
       & 338, 339, 342, 344, 349, 353, 354, 354, 356, 358, 358         &        &        &        &        &         \\
\hline
\end{tabular}}
\caption{Observed p-values for the Rayleigh ($RT$), modified Hermans-Rasson ($HRmT$), Pycke ($PT$), likelihood-ratio $NNTS2$ with $M=1$ ($M=1$) and likelihood-ratio
$NNTS2$ with $M=2$ ($M=2$) tests for the datasets reported by Landler et al. (2021) from the original experiment of Gagliardo et al. (2008) in which they measured the azimuth of vanishing bearings obtained by young homing pigeons randomly assigned to three different groups. The first group (C)
consisted of 41 unmanipulated homing pigeons. The second group (ON) consisted of 27 birds that underwent bilateral olfactory nerve sectioning, and the third
group (V1) included 40 birds that underwent bilateral sectioning of the ophthalmic branch of the trigeminal nerve. Landler et al. (2021) considered subsets only of the C and ON groups. The observed value of the $NNTS2$ test statistic is included below its corresponding p-value.
\label{pigeonsdata} }
\end{center}
\end{table}

\end{document}